\documentclass[twocolumn,aps,prb,showpacs]{revtex4-2}
\usepackage{graphicx,psfrag}
\usepackage{graphics,psfrag}
\usepackage{amssymb,amsfonts,amsmath}
\usepackage{bbm}
\usepackage{dcolumn}
\usepackage{color}
\usepackage{subfig}
\usepackage{rotating}
\usepackage{verbatim}
\usepackage[toc,page]{appendix}

\usepackage{tikz}
 \usepackage{multirow}
 \usepackage{multirow}

\def\ac{{\alpha\gamma}}

\def\kB{{k_{\rm B}}}

\def\rmd{{\rm d}}

\begin{document}

\title{First-principles study of the order-disorder transition in the AlCrTiV high entropy alloy}
\author{Michael Widom}
\affiliation{Department of Physics, Carnegie Mellon University, Pittsburgh PA 15213}

\date{\today}

\begin{abstract}
The AlCrTiV high entropy alloy undergoes an order-disorder transition from body centered cubic (Strukturbericht A2) at high temperatures to the CsCl structure (B2) at intermediate temperatures. We model this transition using first principles Monte Carlo/molecular dynamics simulations. Simulation results yield the temperature-dependent energy, entropy, heat capacity, occupancy fluctuations, and diffraction patterns. The contribution of chemical disorder to the entropy is calculated on the basis of point and pair cluster frequencies. The simulated structures exhibit compensated ferrimagnetism, and the Fermi level lies in a pseudogap. Sensitivity of structure and magnetism to the exchange-correlation functional is discussed, and neutron diffraction experiments are proposed to help resolve the true chemical order.
  
\end{abstract}


\maketitle

\section{Introduction}

Equiatomic AlCrTiV~\cite{Qiu2017} is a high entropy alloy that forms a body centered cubic structure (Strukturbericht A2, Pearson type cI2) from the melt at high temperatures but transforms at intermediate temperatures~\cite{Huang2022a} to a chemically ordered CsCl-prototype structure (B2, cP2). This high entropy alloy is interesting from a practical perspective because of its low density and single phase microstructure, with a high melting temperature and high hardness.  Prior to its consideration as a high entropy alloy, it had been proposed as a possible spin-filter alloy~\cite{Galanakis_2014,Stephen2016,Venkat2018,Stephen2019JAP,Stephen2019PRB} owing to its magnetic semiconducting properties.

From a theoretical perspective, AlCrTiV is interesting both for its electronic and magnetic properties, and also because it evolves from a state of high disorder to one of intermediate disorder. Two of the four chemical species preferentially segregate to one simple cubic sublattice, while the remaining two elements favor the other~\cite{Tian,CA-CPA}. Hints of quaternary Heusler Y-type (prototype LiMgPdSn) ordering have also been reported~\cite{Stephen2016,Venkat2018}. Although long-range chemical order arises, the chemical species on each sublattice remain randomly distributed. Partial order of this type has been proposed as a way to enhance the ductility of B2-type structures~\cite{JieQi} by moderating the strength of the B2 order.

In this paper we apply first principles hybrid Monte Carlo/molecular dynamics simulations~\cite{Widom13} with replica exchange~\cite{Swendsen86} to generate equilibrium distributions of chemical species within the disordered body centered cubic A2 and chemically ordered B2 phases.  Temperature-dependent histograms of calculated energies are analyzed to obtain the energy, entropy, heat capacity and Helmholtz free energy as continuous functions of temperature~\cite{Ferrenberg89} from 700K up to melting around 2000K. Site occupation and nearest-neighbor pair frequencies allow us to calculate the entropy of chemical disorder separately from the total entropy, which includes vibrational, electronic, and magnetic entropy in addition to chemical entropy~\cite{WidomHEABook,Widom16,ADKim}. After our discussion of structure and thermodynamics, we discuss the electronic and magnetic properties of our simulated structures.

\begin{figure}[h!]
  \includegraphics[trim=120mm 5mm 120mm 5mm, clip, width=.4\textwidth]{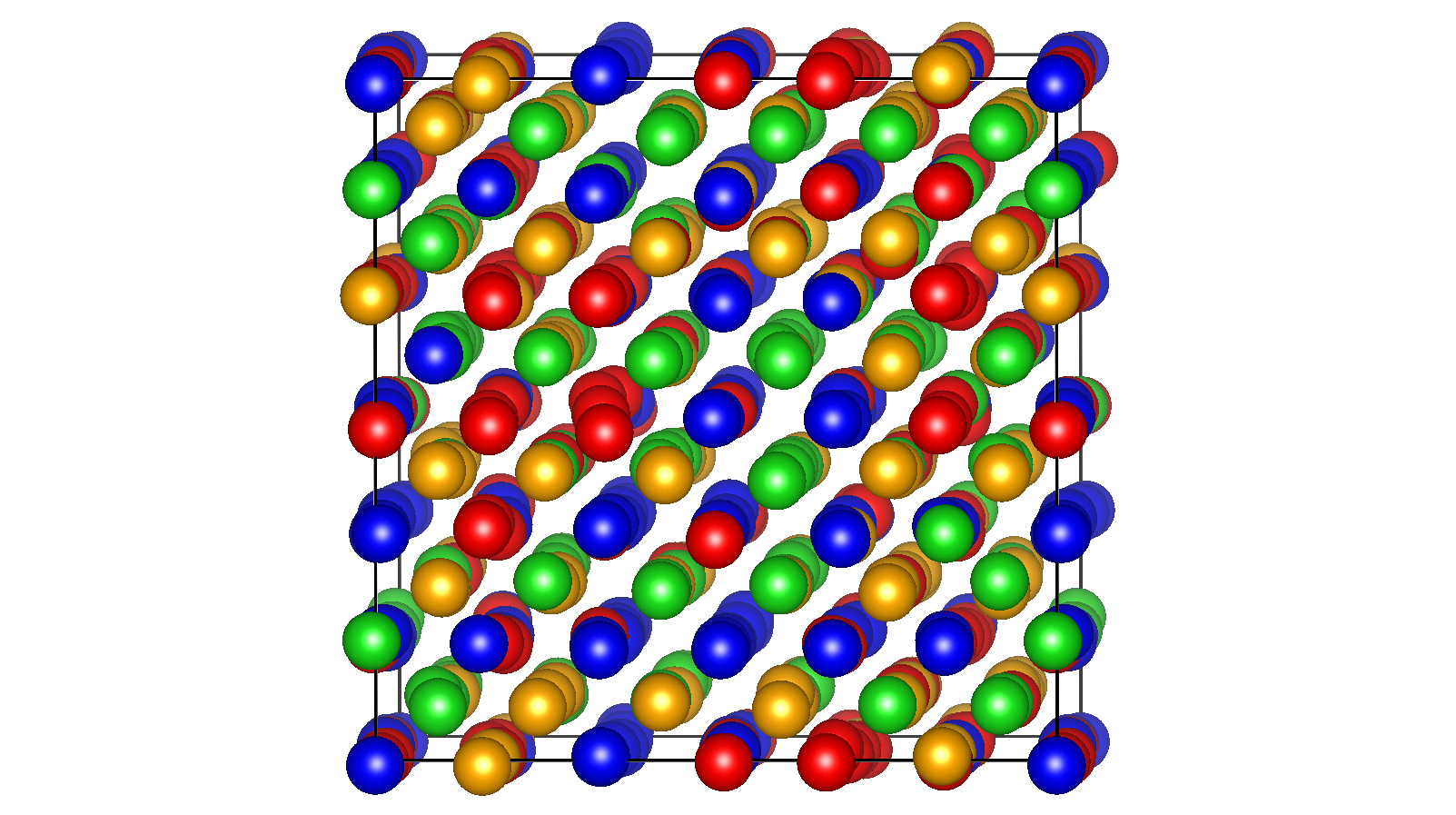}
  \includegraphics[trim=120mm 5mm 120mm 5mm, clip, width=.4\textwidth]{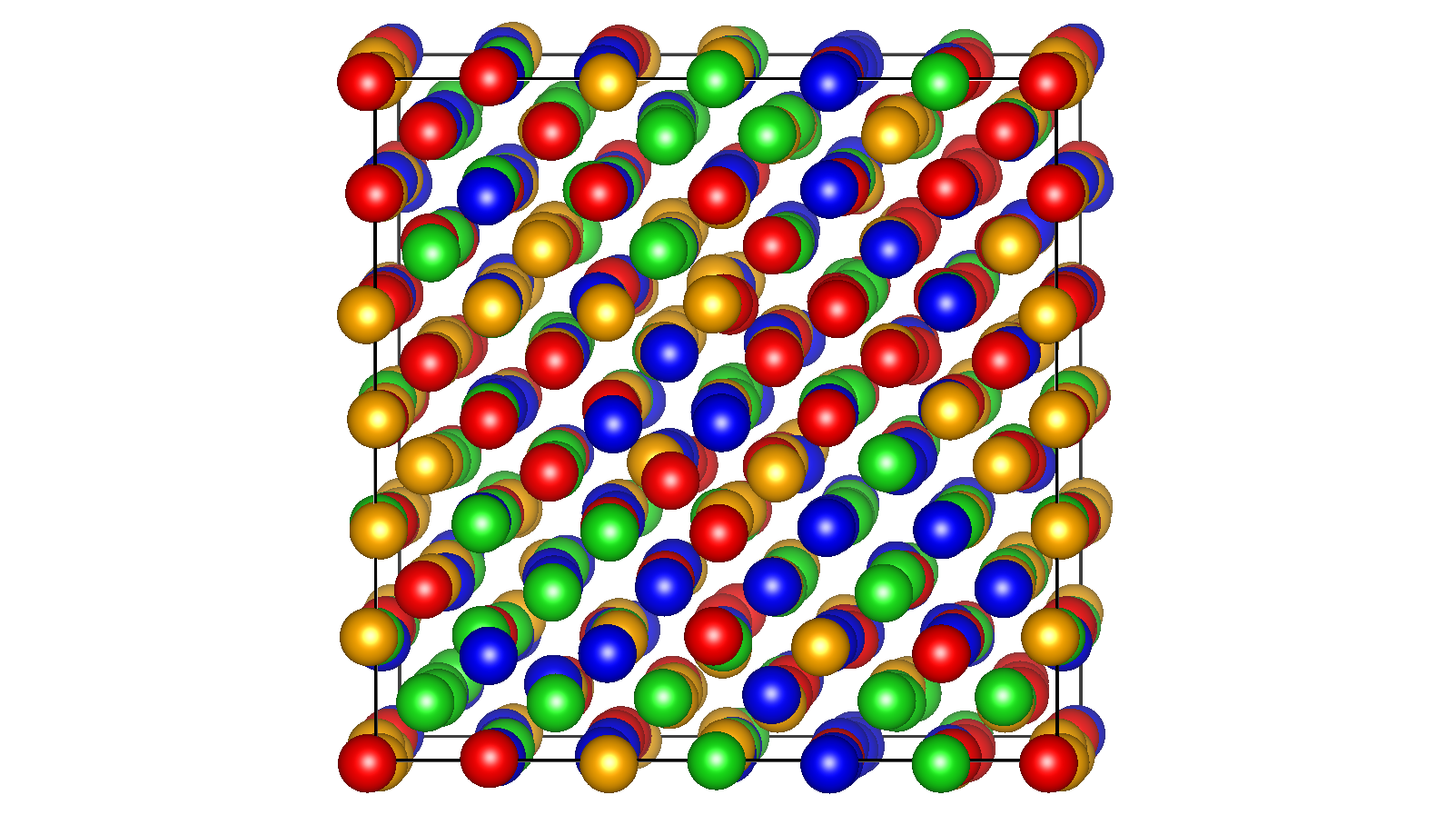}
  \caption{\label{fig:struct} Simulated structures as quenched from T=700K to 300K (top) and quenched from 2000K to 300K (bottom). Chemical species are color coded as Al (green), Cr (orange), Ti (blue), and V (red). Cell size is $6\times 6\times 6$ (432 atoms).}
\end{figure}

Fig.~\ref{fig:struct} illustrates the simulated structures quenched  to room temperature from 700K (below T$_c$) and from 2000K (above T$_c$). Note the alternation of predominantly AlCr (green/orange) layers with predominantly TiV blue/red) layers revealing the B2 order when quenched from 700K. At 2000K the long-range order is absent although some short-range order persists. Based on heat capacity and fluctuations of a chemical long-range order parameter, discussed in Sect.~\ref{sec:methods}, we find that an order-disorder transition occurs in our simulation around 1200-1250K, which is close to the experimentally observed 966C (1239K) for the equiatomic alloy~\cite{Huang2022a} (the experimental melting point is 1805C, or 2078K~\cite{Huang2022b}). In contrast to suggestions that the order at lower temperatures might develop further into a quaternary Heusler type~\cite{Tian,Venkat2018,Hoffman2021}, we find that the true ground state structure is a phase separated mixture of BCC-based chemically ordered binary alloys. This prediction was found to be sensitive to the choice of exchange-correlation functional, as discussed in the appendix.

Through our statistics on site occupation and correlations, we estimate an entropy loss of approximately 1.5$\kB$ due to chemical ordering. We further analyze the structure to predict the diffraction patterns at various temperatures and with different functionals, and propose neutron diffraction as a promising route to detect the chemical order. Finally, we evaluate electronic and magnetic properties, showing the presence of nearly compensated antiferomagnetism~\cite{Galanakis_2014} and a deep pseudogap in the electronic density of states.

\section{Details of calculation}
\label{sec:methods}

Our first principles molecular dynamics (MD) simulations utilize the electronic density functional theory code VASP with projector augmented wave potentials in the PBE~\cite{Perdew96,Kresse99} generalized gradient approximation (GGA). Blocked Davidson iteration was used to solve the Kohn-Sham equations. Equiatomic $L\times L\times L$ supercells were employed with 16 atoms ($L=2$) up to 432 atoms ($L=6$) based on the conventional BCC unit cell with a lattice constant of 3.06~\AA~\cite{Huang2021}. The default plane wave energy cutoff of 240.03 eV was accepted. Monkhorst-Pack $3\times 3\times 3$ and $2\times 2\times 2$ electronic $k$-point meshes were used for $L=2$ and $4$, while only the $\Gamma$-point was used for $L=6$. Cell sizes $L=2$ and $4$ employed spin polarization with assumed initial moments of $+3$ for Cr and $-3$ for V; $L=6$ simulations were nonmagnetic. Additional studies of total energies, and of the electronic and magnetic structure, were converged in $k$-point density and were fully relaxed.

\begin{figure}[h!]
  \includegraphics[width=.48\textwidth]{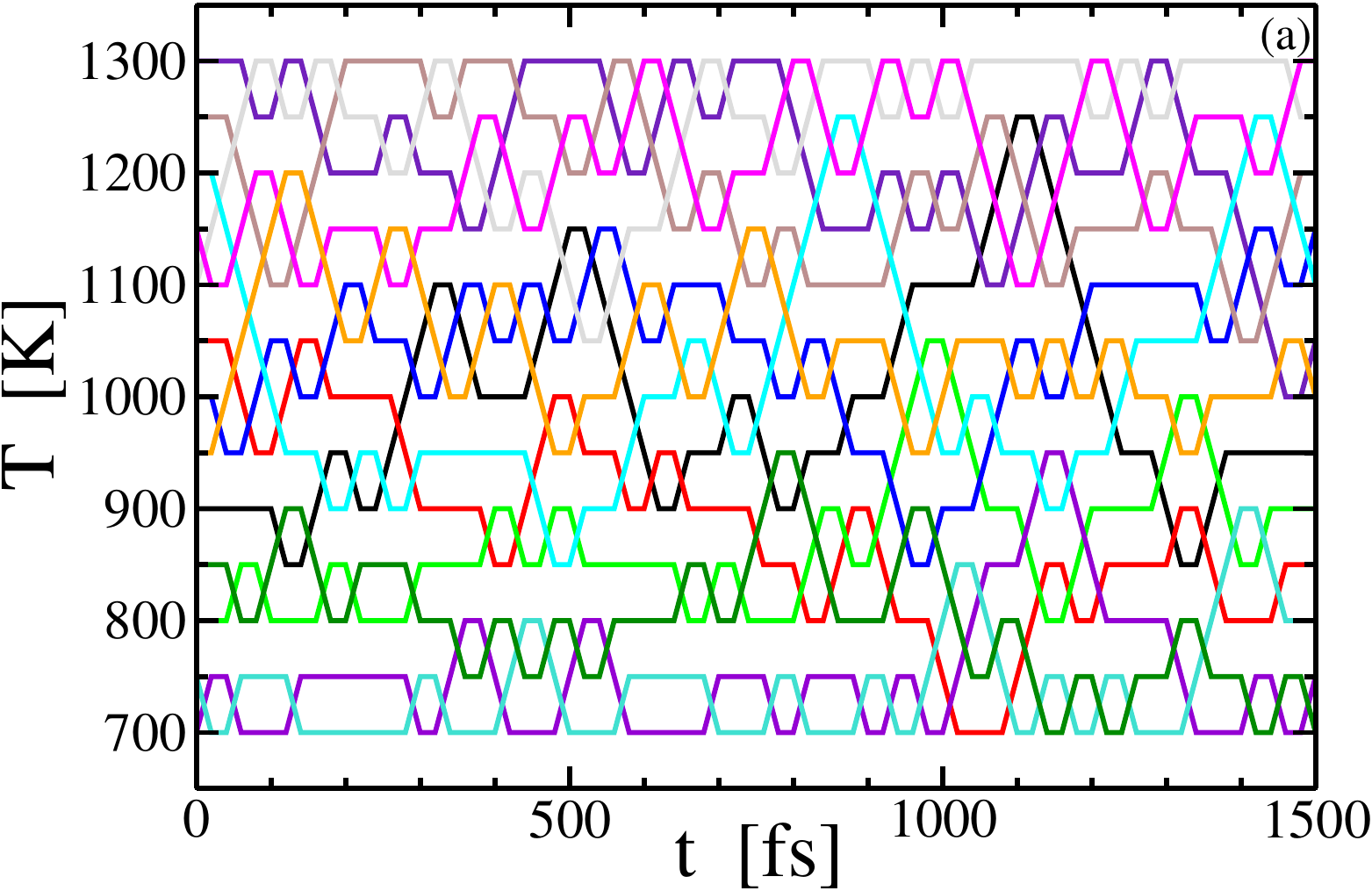}
  \includegraphics[width=.48\textwidth]{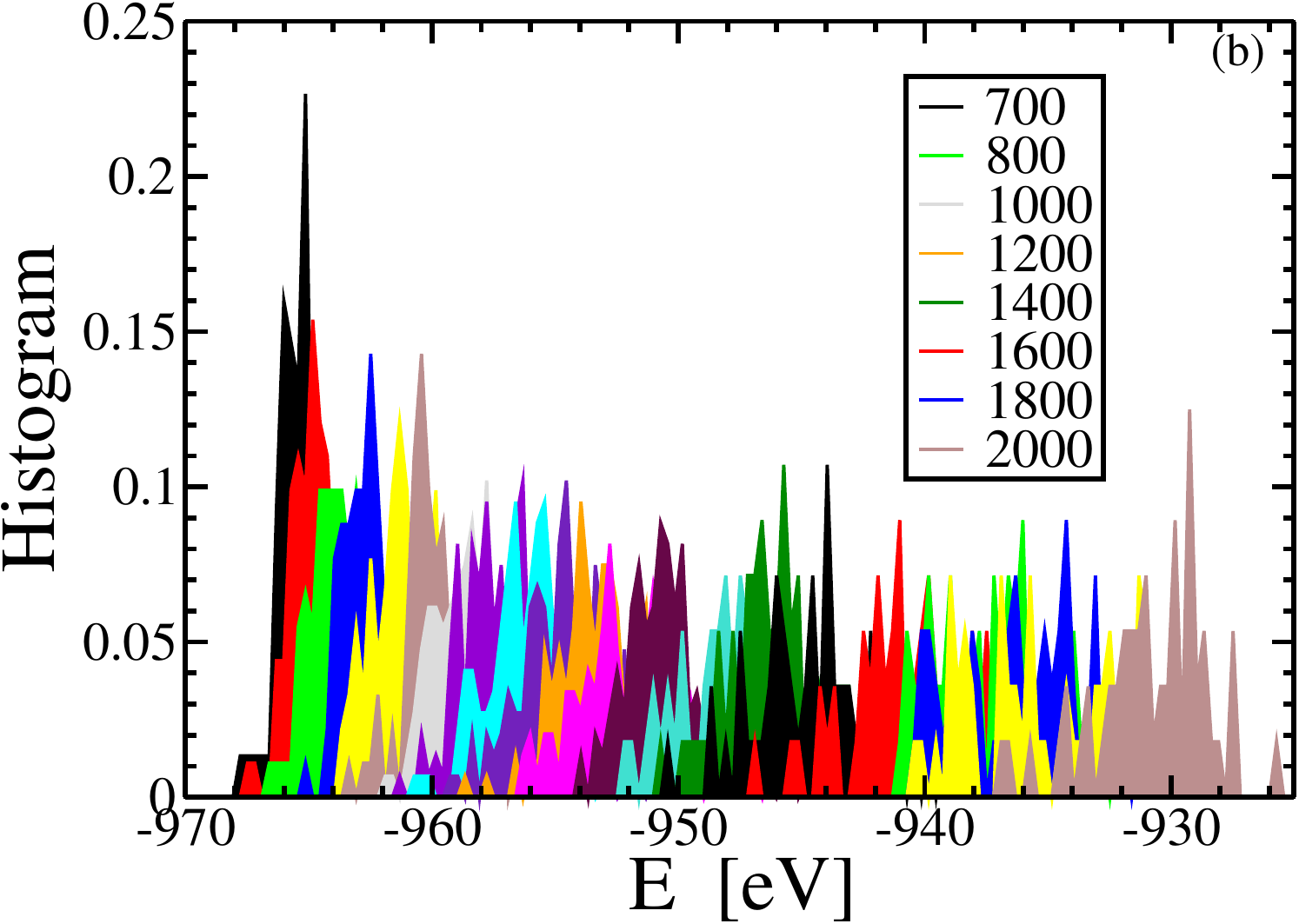}
  \includegraphics[width=.48\textwidth]{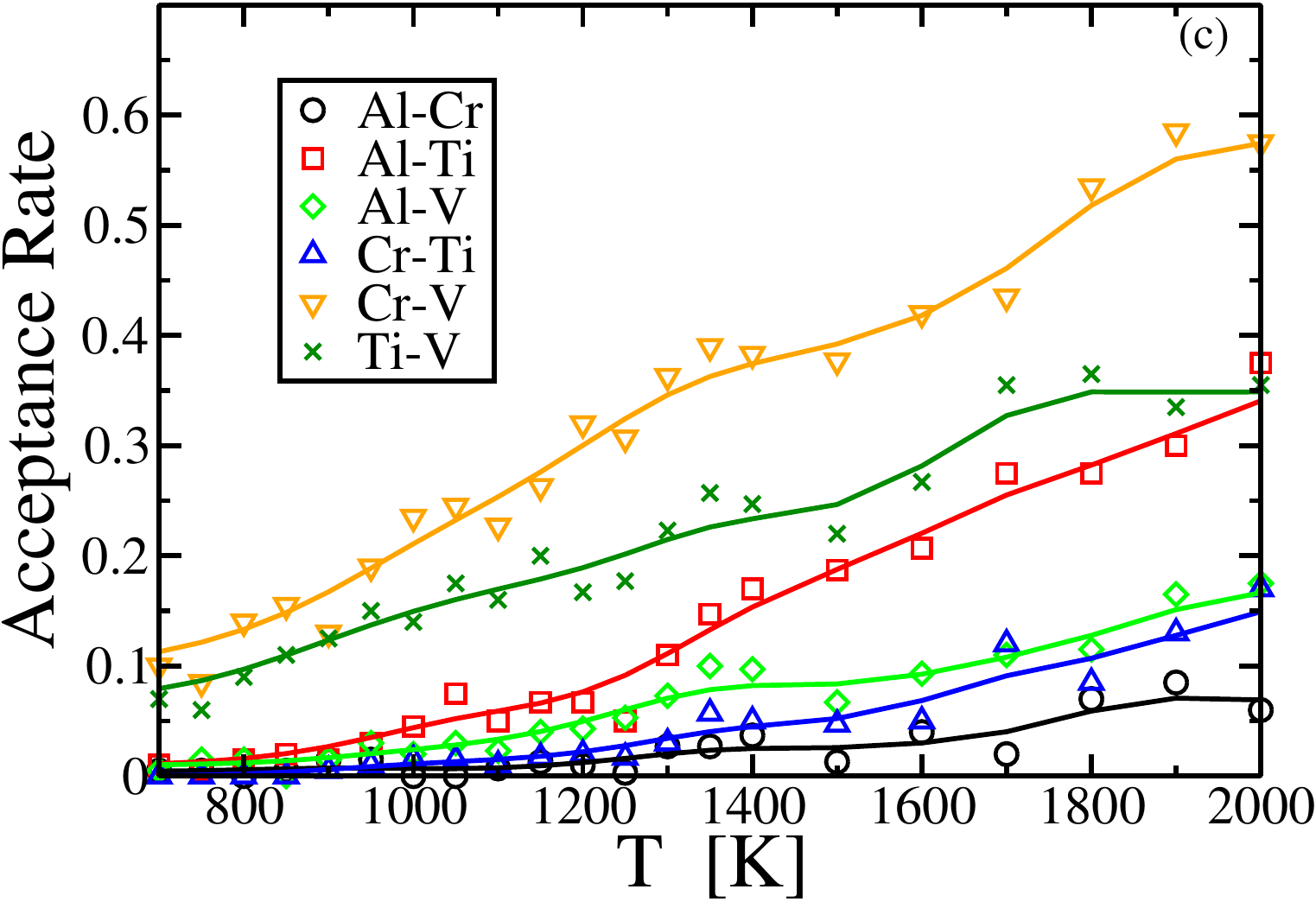}
  \caption{\label{fig:REMCMD} (a) Time-series of temperatures under replica exchange. Independent trajectories are color coded; (b) histograms of energies $H_T(E)$ colored according to equilibrium temperature; (c) chemical species swap rates {\em vs.} temperature. Gaussian smearing of 100K has been applied to interpolate smooth curves; }
\end{figure}

Hybrid Monte Carlo/molecular dynamics methods as described in~\cite{Widom13} alternated 10 attempted Monte Carlo species swaps with 40fs of molecular dynamics using 1fs time steps and velocity rescaling every step. To accelerate the sampling of the configuration space, we supplement the MC/MD with replica exchange as described in~\cite{Swendsen86,ADKim} and illustrated in Fig.~\ref{fig:REMCMD}a. The temperature swap acceptance probability is $\exp{(\Delta\beta\Delta E)}$, where $\Delta E$ is the difference in potential energy between configurations and $\Delta\beta$ is the difference in $1/\kB T$. Acceptance rates for temperature swaps depend upon strong overlap of their energy probability distributions, as illustrated in Fig.~\ref{fig:REMCMD}b. Acceptance rates of Monte Carlo chemical species swaps varied with temperature as illustrated in Fig.~\ref{fig:REMCMD}c.

Given the simulated configurations and their energies, we analyzed their thermodynamic and structural properties taking advantage of histogram methods to interpolate continuous functions of temperature as discussed in~\cite{ADKim,Ferrenberg88,Ferrenberg89}. Specifically, we construct a multidimensional histogram $H(E,M)$, where $E$ is energy and $M$ is any other parameter of interest, then convert to a multidimensional density of states
\begin{equation}
  \label{eq:Omega}
  \Omega(E,M)=\frac{\sum_TH_T(E,M)}{\sum_T e^{(F(T)-E)/\kB T}}
\end{equation}
where the free energies $F(T)$ must be determined self-consistently {\em via}
\begin{equation}
  \label{eq:F}
  \begin{aligned}
    F(T) &=-\kB T\ln{Z(T)},\\
    Z(T) &=\int\rmd E~\rmd M~\Omega(E,M)~e^{-E/\kB T}.
  \end{aligned}
\end{equation}

Internal energy, heat capacity, and entropy may be obtained through analytic derivatives of the free energy $F(T)$. Thermal averages of the the energy $E$ and the order parameter $M$ are obtained as
\begin{equation}
  \label{eq:Mbar}
  \begin{aligned}
  E(T) &=\frac{1}{Z(T)} \int\rmd E~\rmd M~\Omega(E,M)~E~e^{-E/\kB T} \\
  M(T) &=\frac{1}{Z(T)} \int\rmd E~\rmd M~\Omega(E,M)~M~e^{-E/\kB T}.
  \end{aligned}
\end{equation}
Fluctuations (second moments) of the energy yield the heat capacity, $c_v(T)$, and fluctuations of the order parameter yield a generalized susceptibility $\chi_M(T)$, each of which can be thought of as second derivatives of the free energy with respect to a conjugate field.

To obtain single-site occupation statistics $x^{(e,o)}_\alpha$ for species $\alpha$, we mapped the atoms to their nearest lattice sites. To account for the spontaneous symmetry breaking of species to cube vertex (even parity) and cube body center (odd parity) sites, we applied a shift so that the majority of Ti atoms would be assigned to odd parity sites.

The single site occupation provides an immediate approximation for the chemical configurational entropy
\begin{equation}
  \label{eq:S_site}
  S_{\rm site}/\kB = \frac{1}{2}\left\{-\sum_\alpha x^{(e)}_\alpha\ln x^{(e)}_\alpha-\sum_\gamma x^{(o)}_\gamma\ln x^{(o)}_\gamma\right\}.
\end{equation}
We further improve our estimate by correcting for the entropy loss due to nearest-neighbor pair correlations~\cite{Ackermann89,Hoffman2021,ADKim}. This is achieved by calculating the pair frequencies $y^{(eo)}_\ac$ for species $\alpha$ on even sites and $\gamma$ on neighboring odd sites. The mutual information of this pair correlation is
\begin{equation}
  \label{eq:Info}
  I(y^{(eo)}_\ac; x^{(e)}_\alpha x^{(o)}_\gamma) = \sum_\ac y^{(eo)}_\ac  \ln{\left(y^{(eo)}_\ac/x^{(e)}_\alpha x^{(o)}_\gamma\right)}.
\end{equation}
This information must be subtracted from the single site entropy,
\begin{equation}
  \label{eq:S_pair}
  S_{\rm pair} = S_{\rm site} - \frac{z}{2}I,
\end{equation}
where $z=8$ is the coordination number so that $z/2$ is the number of pairs per site.

\section{Structure description}
\label{sec:struct}

\subsection{Model structures}
\label{sec:model}

\begin{figure}[h!]
  \includegraphics[trim=30mm 4mm 30mm 0mm, clip, width=.22\textwidth]{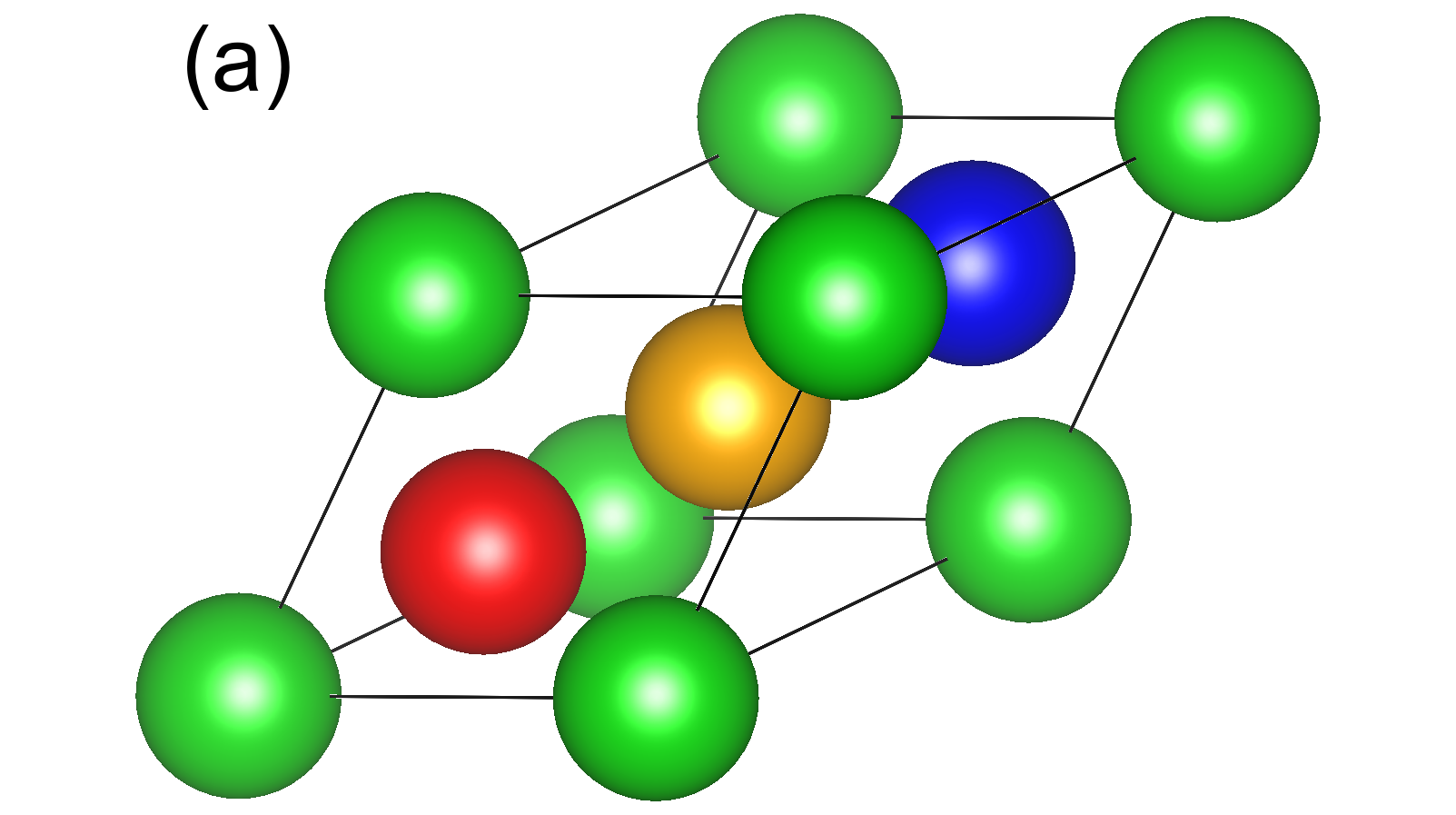}
  \includegraphics[trim=30mm 4mm 30mm 0mm, clip, width=.22\textwidth]{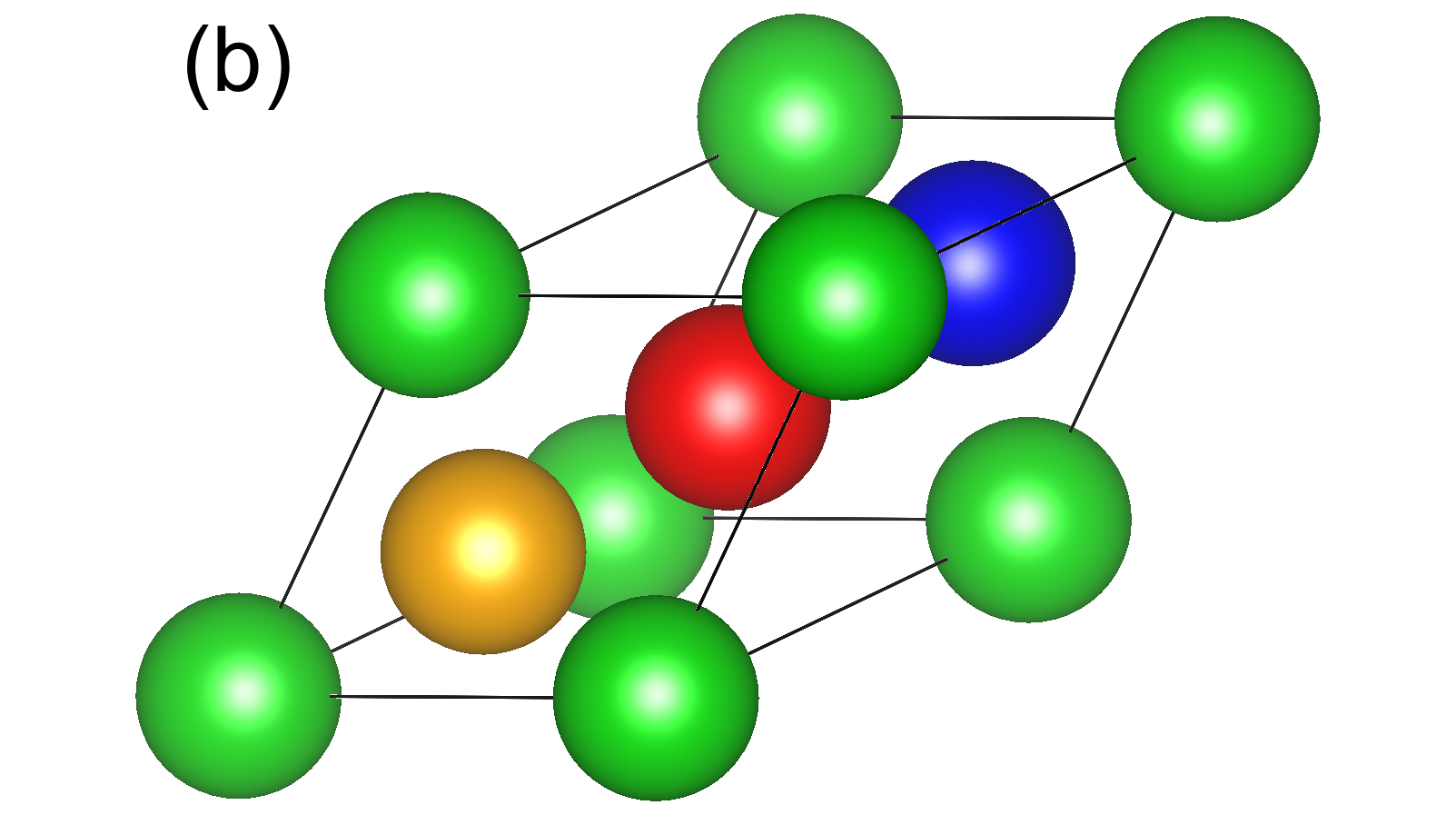}
  \includegraphics[trim=30mm 4mm 30mm 0mm, clip, width=.22\textwidth]{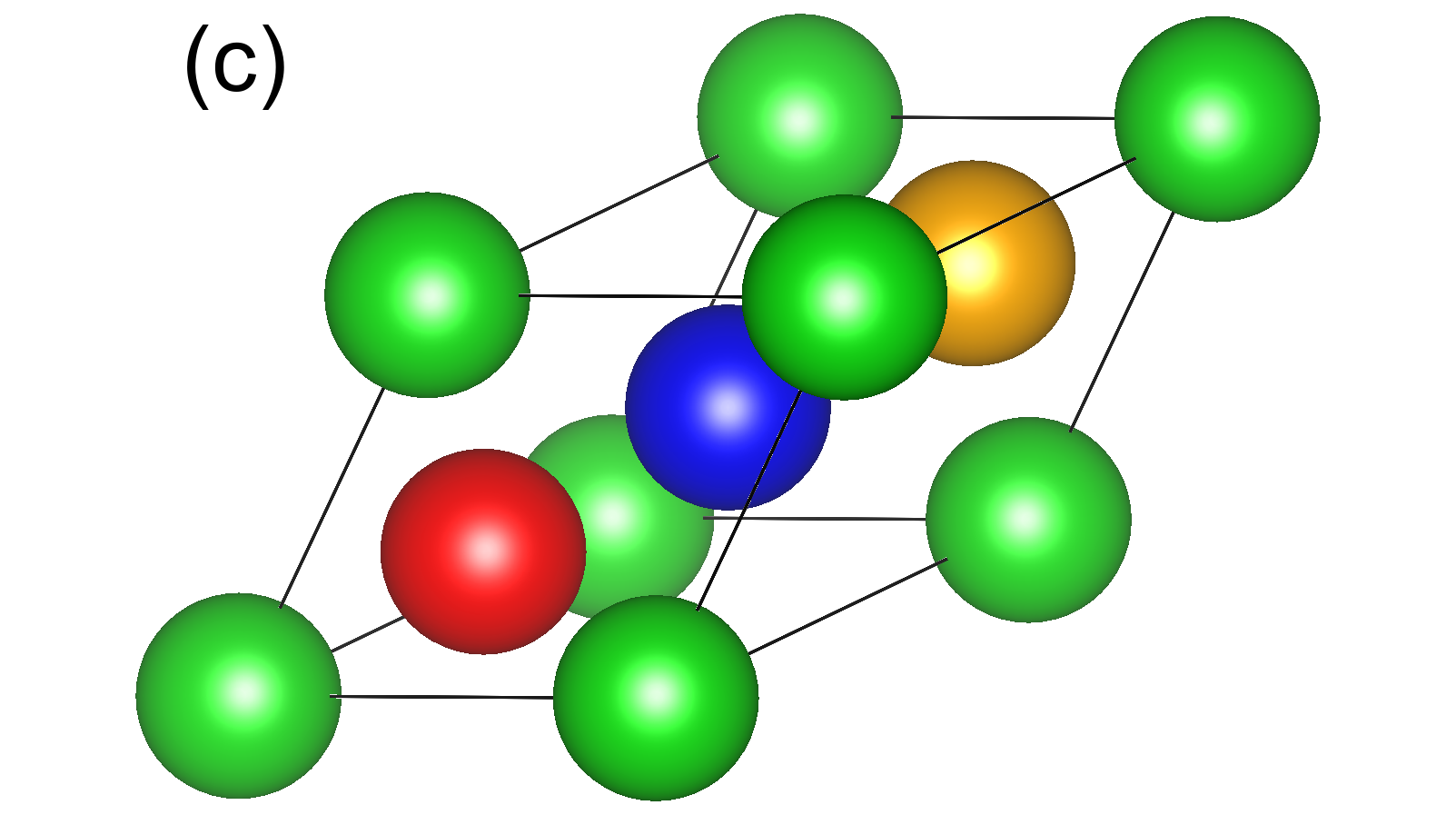}
  \includegraphics[trim=30mm 0mm 40mm 0mm, clip, width=.24\textwidth]{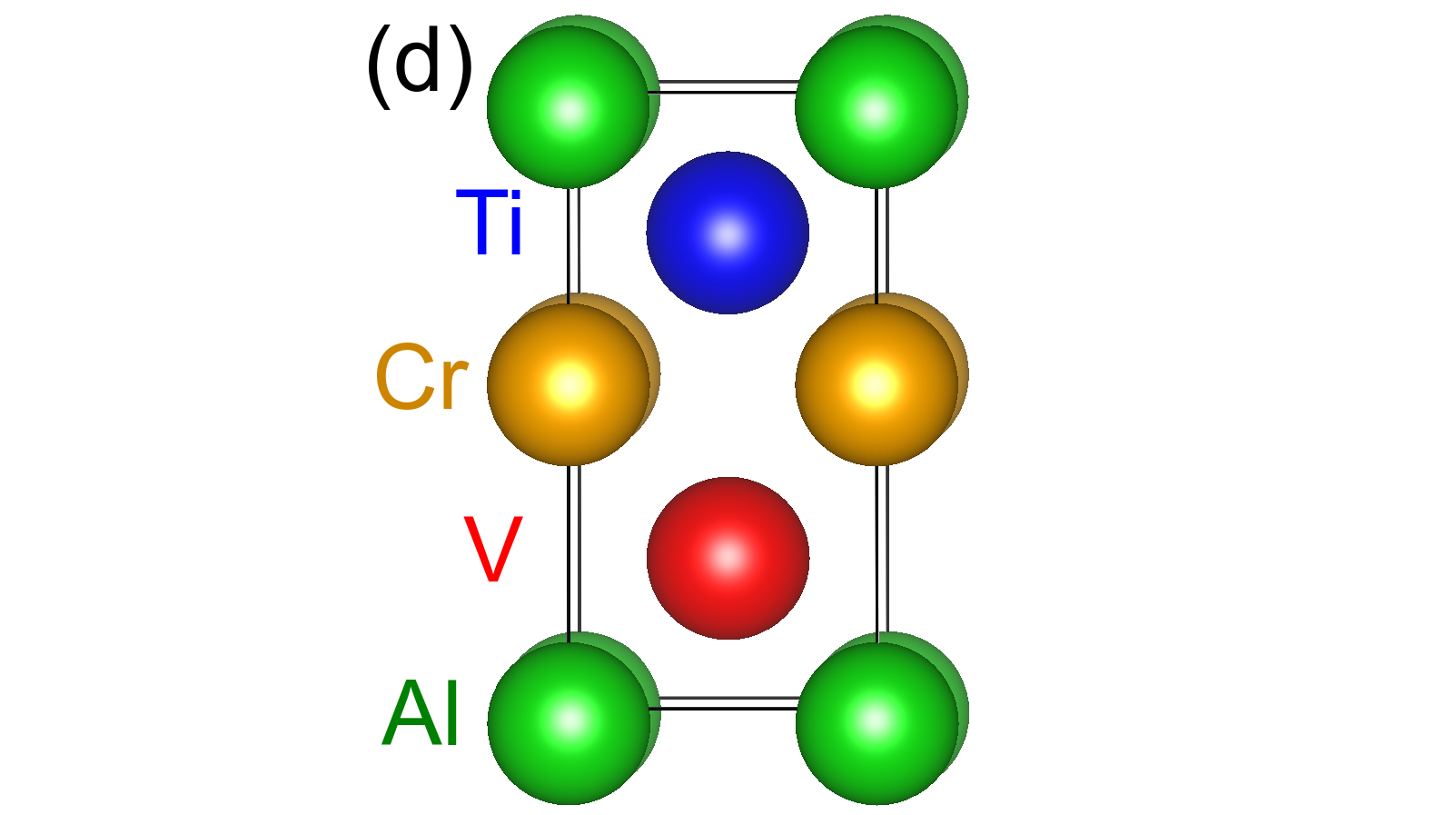}
  \caption{\label{fig:cF16} Primitive cells of the quaternary Heusler structures: (a) Y-I (Al-V-Cr-Ti); (b) Y-II (Al-Cr-V-Ti); (c) Y-III (Al-V-Ti-Cr). The layered tP4 structure (d) is the minimum energy configuration for four atoms. Color coding is Al (green); Cr (orange); Ti (blue); V (red).}
\end{figure}

The quaternary Y-type Heusler structure occupies the 4a, 4b, 4c, and 4d Wyckoff sites of the Pearson type cF16 unit cell with four distinct chemical species. In the four-atom rhombohedral primitive cell of cF16, the sites are arranged along the diagonal in the sequence 4a-4c-4b-4d. By convention~\cite{FelserBook2016} the species are labeled X, X$^\prime$ Y, and Z, with Z a main group element (here Al), and the others arranged in order of decreasing valence. Hence we identify X as Cr, X$^\prime$ as V, and Y as Ti. Owing to the cyclic permutation and inversion symmetries, there are only three distinct assignments of chemical species to crystallographic sites. Following~\cite{Venkat2018,Stephen2019JAP,Stephen2019PRB} we label these structures types Y-I, Y-II, and Y-III, and we specify the sequence of elements along the diagonal, as illustrated in Fig.~\ref{fig:cF16} and listed in Table~\ref{tab:Heusler}.

\begin{table*}
  \begin{tabular}{l|rrrr|l}
    Heusler& \multicolumn{4}{c|}{Wyckoff} & Comments \\
    Type   & 4a & 4c & 4b & 4d & \\
    \hline
    Y-I    & Al &  V & Cr & Ti & nonmagnetic semimetal\\
    Y-II   & Al & Cr &  V & Ti & ferrimagnetic half-semimetal \\
    Y-III  & Al &  V & Ti & Cr & wide-gap ferrimagnetic semiconductor \\
  \end{tabular}
  \caption{\label{tab:Heusler} Three quaternary Heusler structure types. Wyckoff positions and elements are listed in diagonal sequence order: 4a (000); 4c (1/4 1/4 1/4); 4b (1/2 1/2 1/2); 4d (3/4 3/4 3/4). Comments are properties as reported by~\cite{Venkat2018} and~\cite{Stephen2019PRB}, and confirmed by us.}
\end{table*}

We will also consider B2-(ab)-(cd) structures that occupy the 4a and 4b sites equivalently, and likewise the 4c and 4d sites. For example, the model B2-(AlCr)-(TiV) randomly places Al and Cr on the cube vertex sites (even), with Ti and V randomly placed on the body centers (odd). Prior studies employing the coherent potential approximation~\cite{Tian,CA-CPA} have ranked the energies, finding that B2-(AlCr)-(TiV) is most favorable, and B2-(AlTi)-(CrV) is the least favorable. Those prior calculations were performed without spin polarization.

\subsection{Simulated structures}

Fig.~\ref{fig:struct} illustrates simulated structures with $L=6$ (432 atoms). The top part shows a snapshot of an equilibrium structure at T=700K that we quenched to 300K through molecular dynamics without Monte Carlo swaps. Notice the layers and columns of predominantly Al/Cr atoms (green/orange) alternating with layers and columns of Ti/V atoms (blue/red). The distributions of Al/Cr within layers and columns are random, as are the distributions of Ti/V. The order is predominantly of the B2-(AlCr)-(TiV) type.  Nonetheless, the order is not complete. Some (TiV) enter into the predominantly (AlCr) sites and {\em vice-versa}. In contrast, the bottom part shows a snapshot of the structure equilibrated at T=2000K (also quenched to 300K). Although a tendency to alternate (AlCr) with (TiV) remains, it is less than was present in equilibrium at 700K, and the even/odd site distinction decays with increasing separation.

\begin{figure}[h!]
  \includegraphics[width=.48\textwidth]{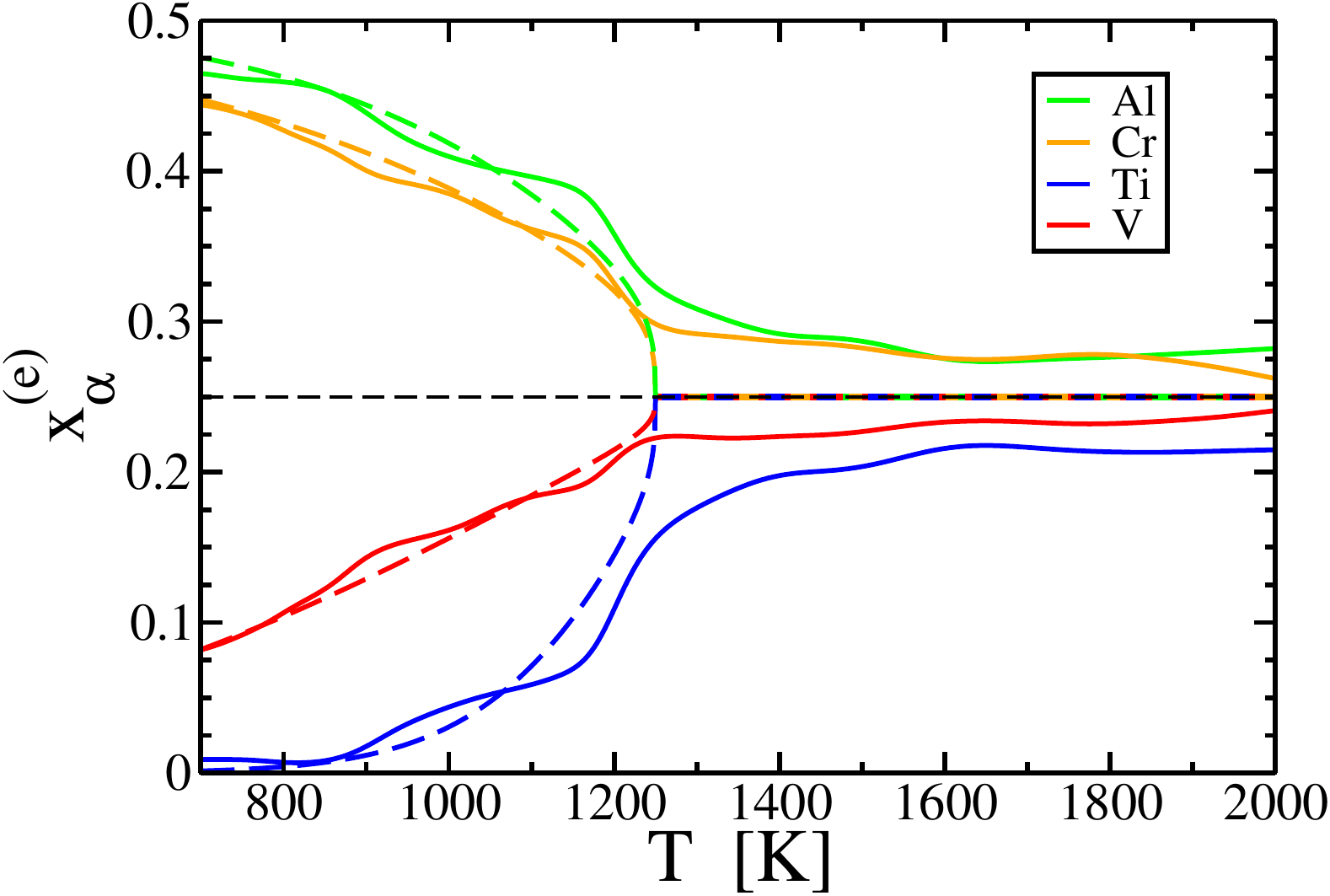}
  \caption{\label{fig:XofT} (solid lines) Temperature-dependent occupation of cube vertex (even) sites; (dashed lines) guide to the eye to presumed thermodynamic limit. Cell size is $6\times 6\times 6$ (432 atoms).}
\end{figure}

\begin{figure}[h!]
  \includegraphics[width=.48\textwidth]{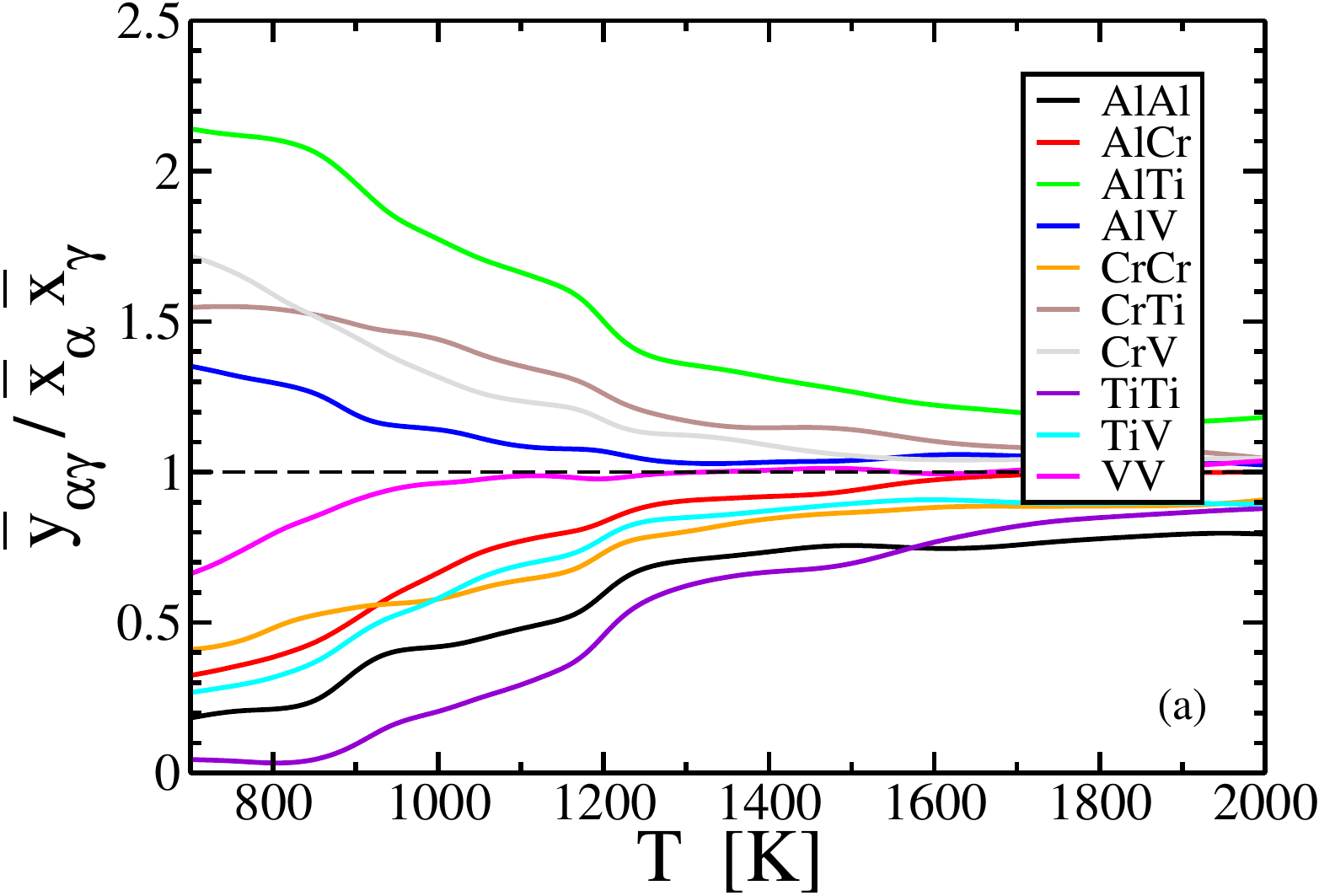}
  \includegraphics[width=.48\textwidth]{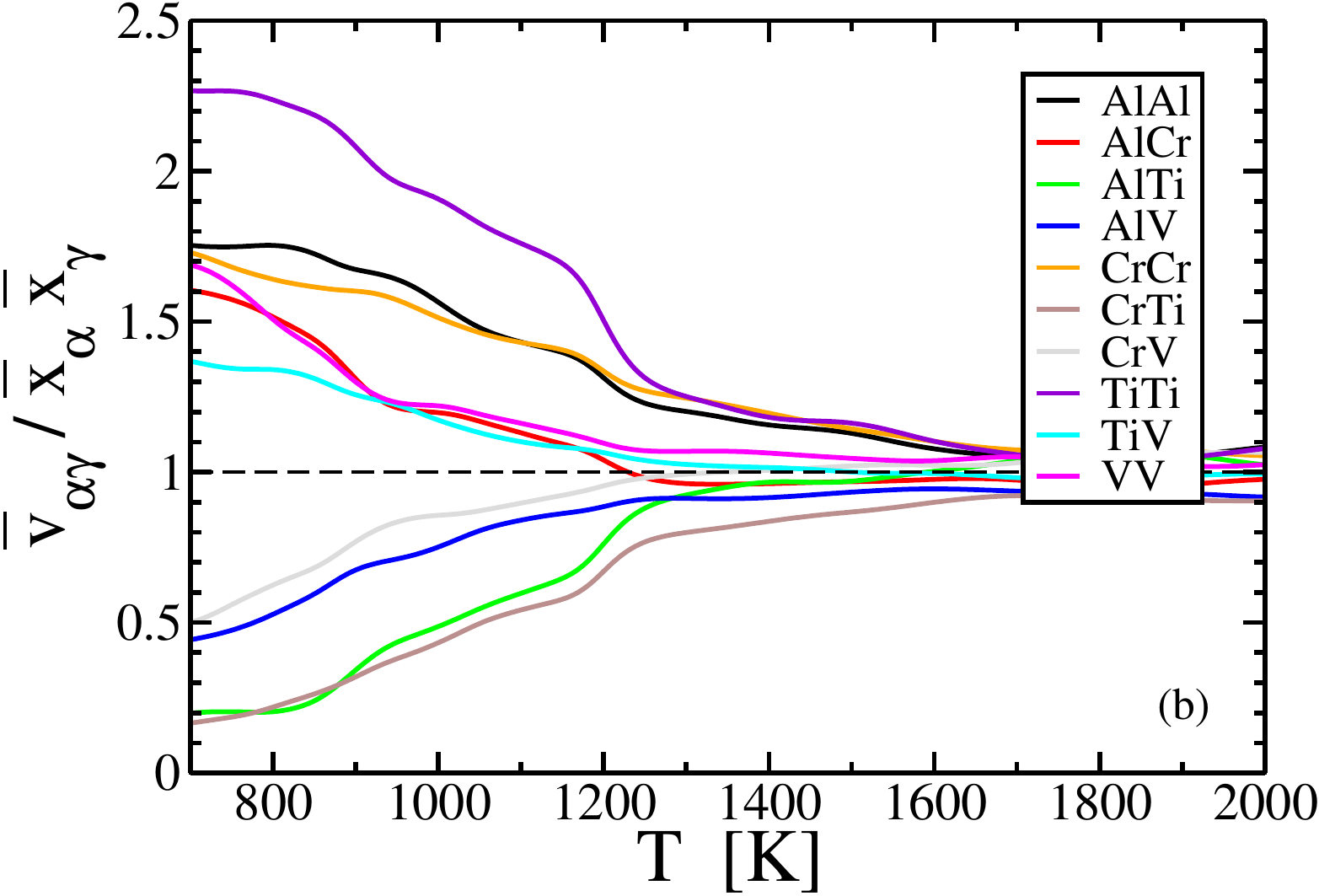}
  \caption{\label{fig:YVofT} Temperature-dependent symmetrized pair frequencies for (a) nearest neighbors and (b) next nearest neighbors. Cell size is $6\times 6\times 6$ (432 atoms).}
\end{figure}

After mapping the atoms to the nearest lattice sites, we can evaluate the point occupations $x^{(e,o)}_\alpha$. their temperature dependences are shown in Fig.~\ref{fig:XofT}. At high temperatures the single site occupancies tend towards uniformity, although because of our mapping to place the majority of Ti on the odd sites, some bias remains that would vanish in the limit of larger simulated system sizes. A transition is evident around $T_c = 1250$K, below which the species segregate to sites of specific parity (experimentally it occurs around 1239K~\cite{Huang2019}). Simulations in small cells are expected to overestimate transition temperatures because they overstate the degree of order.  As a guide to the eye, we plot a presumed thermodynamic limit assuming the deviation of $x^{(e)}_\alpha$ from $1/4$ vanishes as $(T_c-T)^\beta$, taking $\beta=0.326$, which is the universal value for Ising transitions in three dimensions.

The pair frequencies (see Fig.~\ref{fig:YVofT}), in contrast, show significant deviations from ideality even close to the melting point around 2000K. This shows the presence of substantial short-range chemical order even in the disordered A2 phase that lacks genuine long-range order. We have symmetrized the frequencies by defining $\bar{y}_\ac=(y^{(eo)}_\ac+y^{(oe)}_\ac)/2$, and we normalize them by dividing by the ideal mixing expectation $\bar{x}_\alpha\bar{x}_\gamma$. Deviations of pair frequencies from ideal mixing accelerate below $T_c$. Notice that AlTi nearest-neighbor pairs are the most common, with AlAl and especially TiTi being the least. The trend is reversed for next-nearest pairs ($\bar{v}_\ac$), with like species being the most prevalent.

\begin{figure}[h!]
  \includegraphics[width=.48\textwidth]{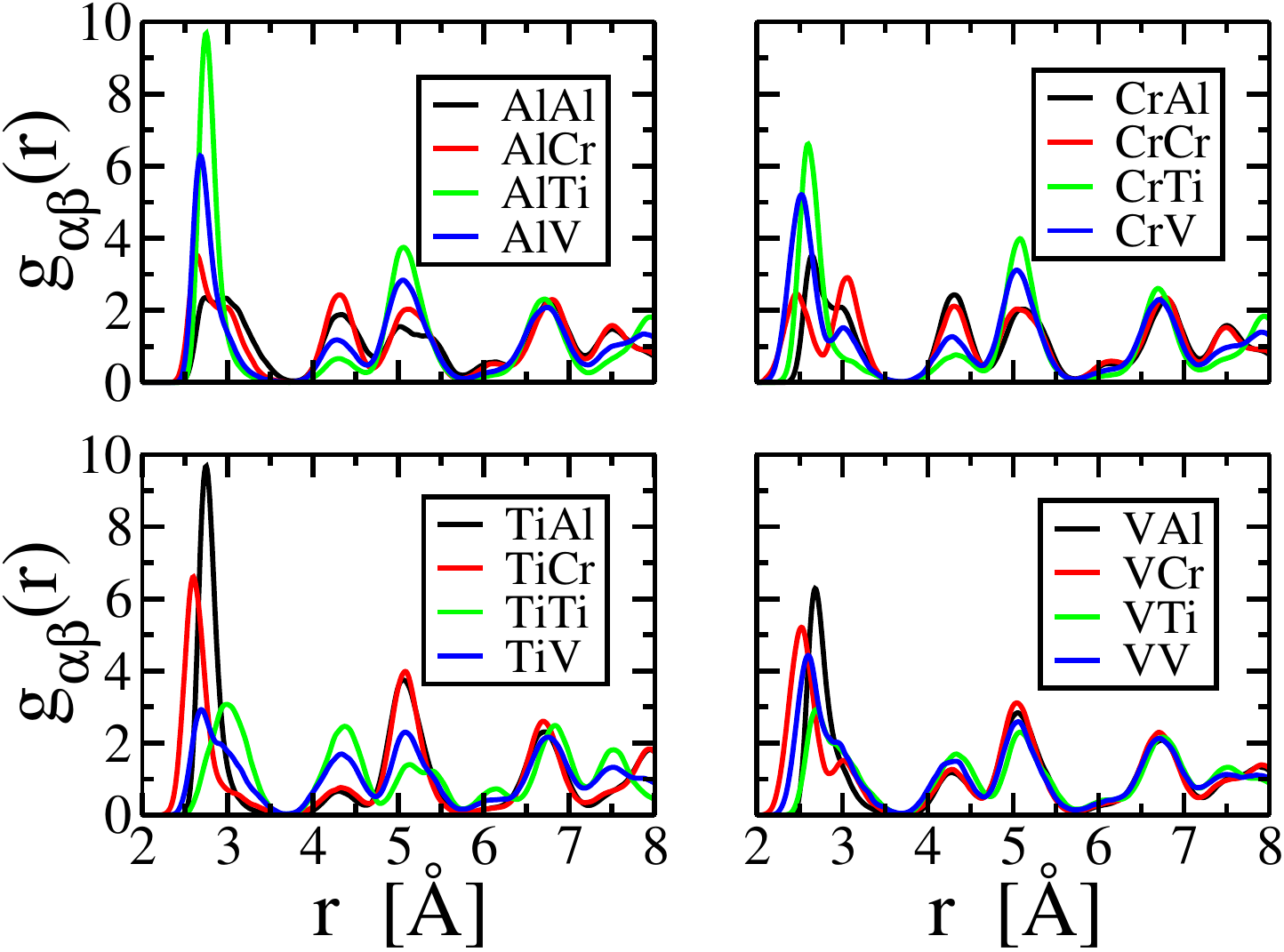}
  \caption{\label{fig:GofR} Partial pair correlation functions $g_\ac(r)$ of our $6\times 6\times 6$ supercell simulation quenched from 1000K to 300K.}
\end{figure}

The structure can be further characterized using partial pair correlation functions, $g_\ac(r)$  (see Fig.~\ref{fig:GofR}), that give the relative probability for pairs of atoms of species $\alpha$ and $\gamma$ to occur at separation $r$. Double-peaked structures around 2.8 and 3.2~\AA~ reveal the nearest and next-nearest separations of body-centered cubic structures. Widths of the peaks exhibit the combined influence of lattice distortion (also visible in Fig.~\ref{fig:struct}) and thermal broadening~\cite{Feng2018}. The $g_{\rm Al\gamma}$ correlations show strong nearest-neighbor peaks for AlTi pairs on sites of opposite parity, while AlAl and AlCr appear at next-nearest neighbors on sites of identical parity, a pattern that continues out to further neighbors. The situation is similar, but weaker, for the Cr$\gamma$ pairs. The Ti$\gamma$ correlations are, likewise, stronger than the V$\gamma$.

Diffraction experiments are commonly employed to reveal crystal structure. Although it is convenient for the identification of B2-like chemical order in transition metal aluminides, X-ray diffraction is rather insensitive to the nature of chemical ordering among the transition metals, owing to a lack of contrast. Indeed, we find the X-ray powder diffraction patterns (not shown) of our simulated structures resemble the report of~\cite{Venkat2018} equally well as their proposed model, which interchanges the roles of V and Cr.

\begin{figure}[h!]
  \includegraphics[width=.48\textwidth]{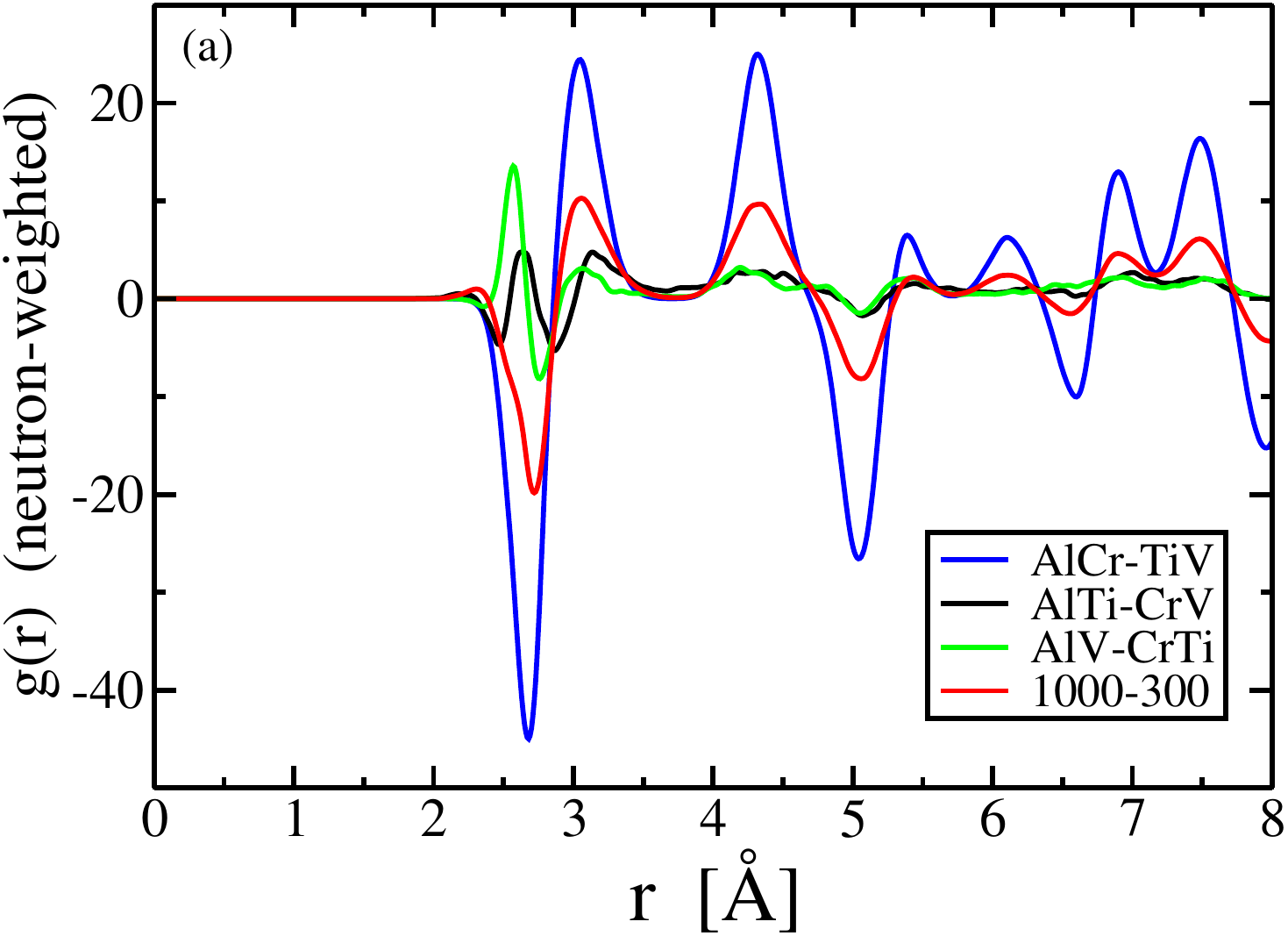}
  \includegraphics[width=.48\textwidth]{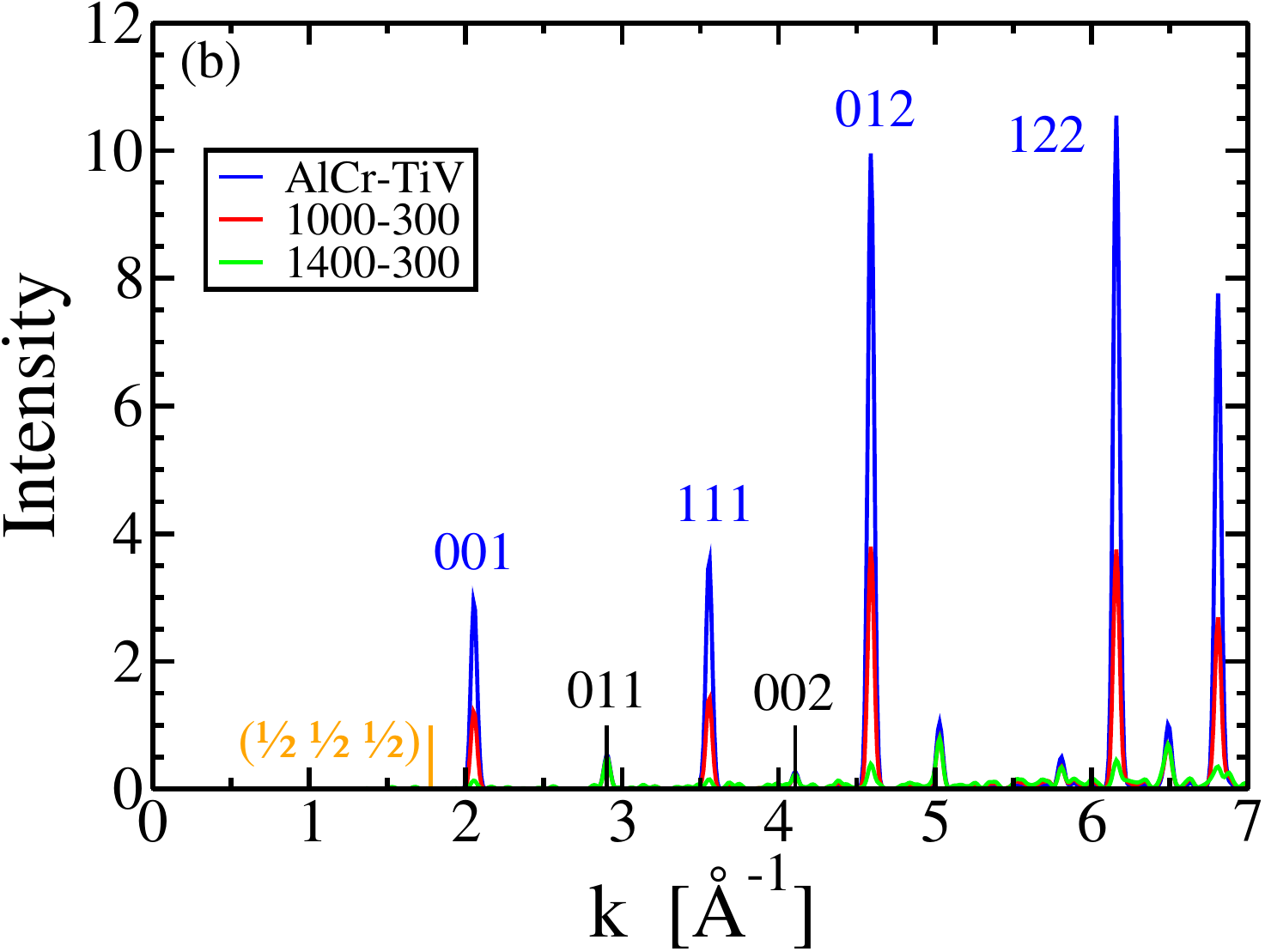}
  \caption{\label{fig:neutron} (a) Neutron-weighted pair correlations of the three B2 variants, and of our $6\times 6\times 6$ supercell simulations quenched from 1000K to 300K; (b) Neutron diffraction patterns of the B2 variant AlCr-(TiV), and of our simulations quenched from 1000K and from 1400K to 300K. Miller indices are marked according to the conventional BCC unit cell with $a=3.06$~\AA. BCC peaks ($h+k+l$ even) in black, B2 peaks ($k+k+l$ odd) in blue, and Heusler (half integer) in orange.}
\end{figure}

The structure might be better probed experimentally using neutron diffraction, because of the differing magnitudes and signs of the neutron scattering lengths, which are $b^{\rm coh}=3.449, -3.438, -0.3824, 3.635$, respectively, for Al, Ti, V, and Cr. Fig.~\ref{fig:neutron}(a) shows that the neutron-weighted pair correlation functions differ markedly between various types of B2 order due to the contrasting neutron scattering lengths.  Magnetic scattering effects are neglected here, but they would further enhance the contrast of Cr with V. Our simulated structure is a good match to B2-(AlCr)-(TiV) in terms of signs and relative magnitudes of peaks, but has lower amplitude because the segregation of elements to sublattice in our simulation is not complete.

B2-type order at 700K is clearly demonstrated in the diffraction pattern (Fig.~\ref{fig:neutron}(b)) through the presence of peaks with Miller indices with $h+k+l$ odd that are extinct in the BCC structure and also in the 1400K structure. Indeed, these peaks are greater in strength than the BCC peaks with $h+k+l$ even, because of the near cancellation of scattering lengths among the four elements. Diffraction peaks of the Heusler structures such as $L2_1$ and $D0_3$ would occur at half integer Miller indices because their cF16 cells are $2\times 2\times 2$ supercells of BCC (cI2); they are completely absent in our simulation.

To test the impact of spin polarization on our replica exchange MC/MD results we re-ran our $4\times 4\times 4$ supercells across the temperature range 1000-1800K. It was helpful to employ the preconditioned conjugate gradient algorithm to simultaneously update the electronic orbitals, and to allow up to 120 electronic steps for convergence; other algorithms proved occasionally unstable. In addition, we found the magnetic moments were highly sensitive to the atomic volume, and we took cell dimensions of 12.8~\AA~ ($4\times a=3.2$) in order to achieve high magnetic moments. No noteworthy impact of magnetism on the structure was observed after 10$^3$ attempted swaps at each temperature.

\section{Thermodynamics}
\label{sec:thermo}

We can understand the preferences for chemical ordering through consideration of the enthalpy of mixing,
\begin{equation}
  \label{eq:DeltaH}
  \Delta H_{\rm For} = E - \sum_\alpha x_\alpha E_\alpha.
\end{equation}
Here $E$ is the total energy per atom of a given compound, and $E_\alpha$ is the energy per atom of species $\alpha$ in its pure state, and $x_\alpha$ is the global concentration of species $\alpha$. The convex hull of the set of $\Delta H_{\rm for}$ values across the composition space identifies the set of thermodynamically stable single phases and phase mixtures. We also define an {\em instability} energy~\cite{Fe-glass}, $\Delta E$, for a given structure as the height of its $\Delta H_{\rm for}$ above the convex hull. Positive $\Delta E$ values indicate a lack of stability  at low temperatures and a tendency to phase separate into a mixture of coexisting phases. Negative $\Delta E$ is defined as the energy relative to the convex hull omitting the structure in question. We find that all equiatomic AlCrTiV quaternaries have positive $\Delta E$ and phase separate into a mixture of AlTi in a Pearson type tP4 structure, together with Cr$_3$V$_4$ and Cr$_4$V$_3$ in Pearson type hR7 structures.

Table~\ref{tab:dE} lists the energies of selected equiatomic structures ($x_\alpha=1/4$), starting from the above-mentioned phase separated state. All compositions are equiatomic, spin polarized, fully relaxed, and employ the PBE exchange correlation functional. The layered tP4 structure illustrated in Fig.~\ref{fig:cF16}, in which Al layers are surrounded by Ti and V layers, achieves the lowest positive $\Delta E$. This is not surprising since, owing to differences in electronegativity, Al binds strongly to transition metals, with the strongest attraction to Ti ($\Delta H=400$ meV/atom in AlTi, Pearson type tP4), followed by V ($\Delta H=293$ meV/atom in Al$_3$V, Pearson type tI8), and finally Cr ($\Delta H=-144$ meV/atom in Al$_{11}$Cr$_4$, Pearson type aP15). Slightly higher in energy are a variety of 128-atom structures drawn from our simulations at 700K and then fully relaxed.

The average energy of ten structures drawn from our $L=4$ simulation at 700K and then relaxed lies slightly above the average energy of B2-(AlCr)-(TiV) structures. They lie significantly below the average energy of structures relaxed (using PBE) from our r2SCAN simulation (see also Appendix~\ref{app:r2scan}). Both the B2-(AlCr)-(TiV) and the relaxed simulated structures lose magnetism during relaxation. The three different quaternary Heusler structures are all higher in energy. We label them by the sequence of species along the 16-atom cube diagonal. The Y-III Heusler has the lowest energy of the three, as expected given its strong ferrimagnetism and wide bandgap.

\begin{table}
  \begin{tabular}{lll}
    $\Delta E$ & Structure & comment \\
    \hline
    0 & Convex hull & phase separated \\ 
    38  & layered tP4 & nonmagnetic metal with pseudogap \\
    67  & B2-(AlCr)-(TiV) & nonmagnetic metal with pseudogap \\
    73  & $L=4$ 700K & simulated using PBE \\
    125 & $L=4$ 700K & simulated using r2SCAN \\
    129 & Y-III Heusler & Al-V-Ti-Cr ferri insulator \\
    146 & B2-(AlV)-(CrTi) & nearly compensated ferrimagnet\\
    159 & Y-II Heusler & Al-Cr-V-Ti half semimetal\\
    205 & B2-(AlTi)-(CrV) & nearly compensated ferrimagnet\\
    214 & Y-I Heusler & Al-V-Cr-Ti nonmagnetic semimetal \\
    \end{tabular}
    \caption{\label{tab:dE} Energies relative to convex hull of various simulated and hypothetical structures. The PBE functional has been used. Units are meV/atom.}
\end{table}

\begin{figure}[h!]
  \includegraphics[width=.48\textwidth]{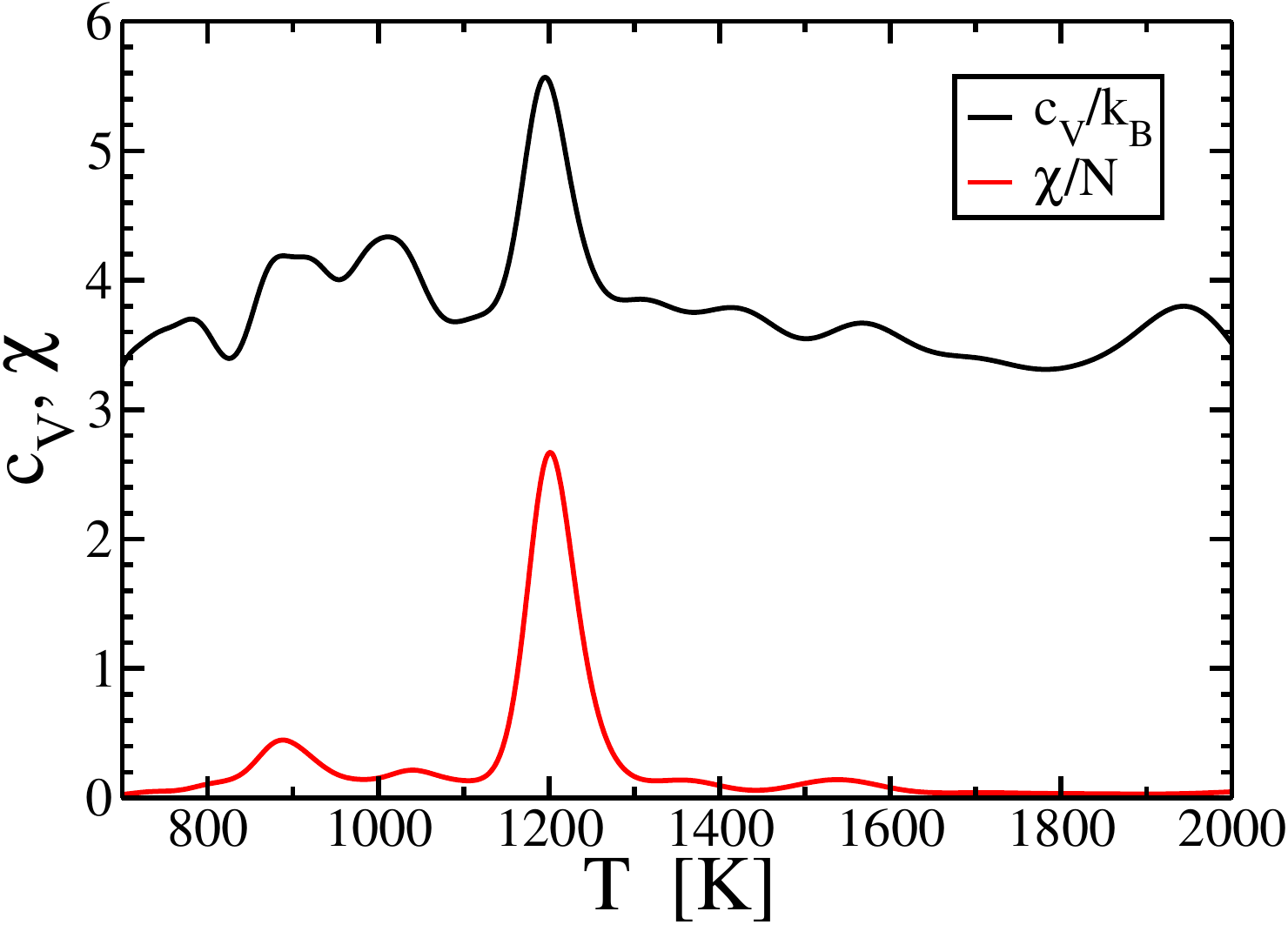}
  \caption{\label{fig:c-chi} Temperature-dependent specific heat capacity and and susceptibility.}
\end{figure}

As discussed in Section~\ref{sec:methods}, our histogram methods~\cite{ADKim} allow us to calculate thermodynamic properties as continuous functions of temperature. In Fig.~\ref{fig:c-chi} we plot the heat capacity $c_v$ and susceptibility $\chi$. The heat capacity is obtained from fluctuations in the energy. Over the range 700-2000K is remains close to the classically expected value of $3\kB$, but exhibits a distinct peak at 1200K, indicative of a phase transition. The susceptibility is obtained from fluctuations of a chemical order parameter $M\equiv x^{(e)}_{Al}-x^{(e)}_{Ti}$. It displays a strong divergence at 1200K, indicating a likely Ising-type continuous transition.

\begin{figure}[h!]
  \includegraphics[width=.48\textwidth]{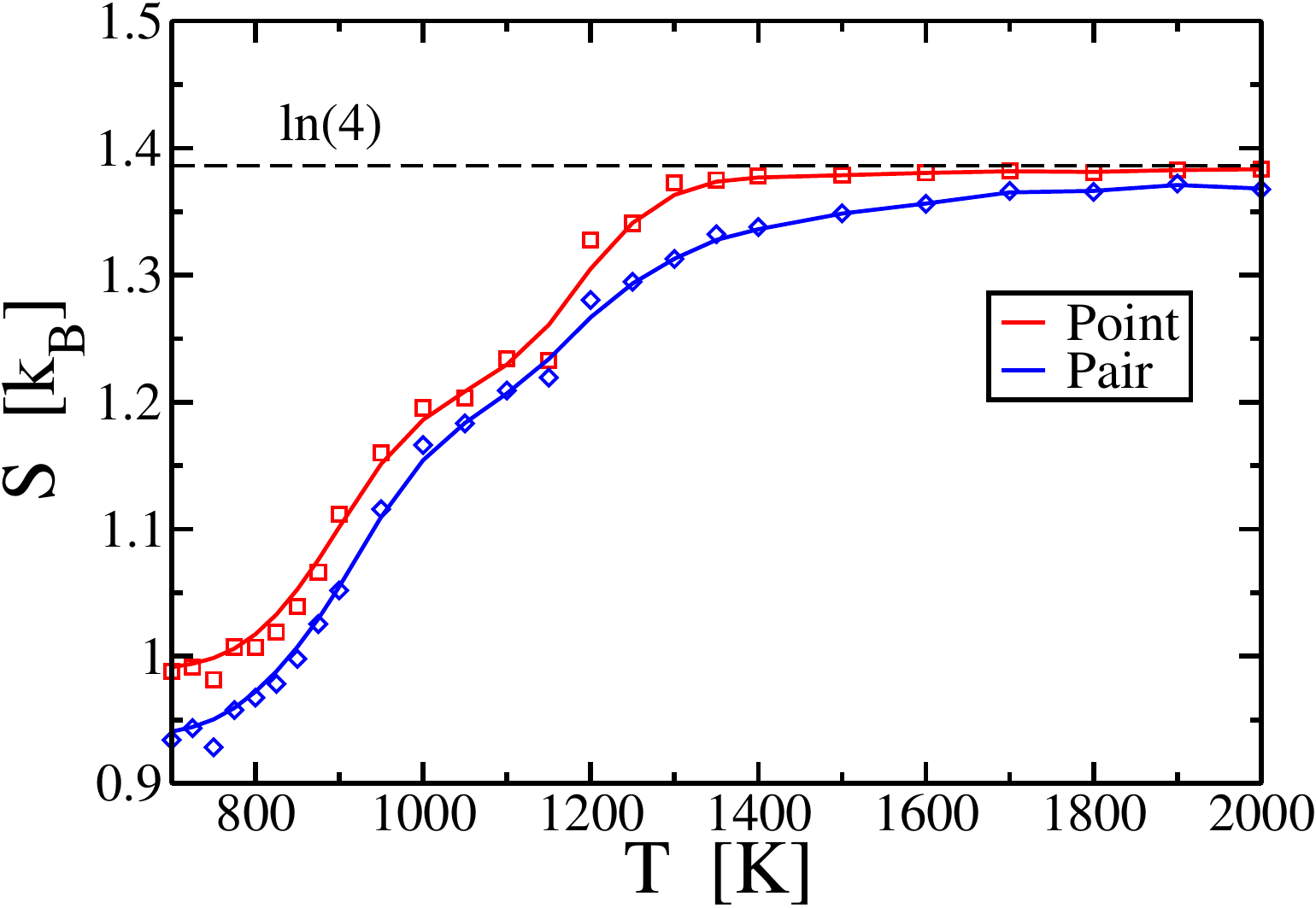}
  \caption{\label{fig:S} Configurational entropy derived from pair and point occupation statistics.}
\end{figure}

The ideal mixing entropy for an equiatomic quaternary is $\ln{4}\kB$. Segregation of species to BCC sublattice will reduce the entropy, and site occupation correlations will reduce it further. Fig.~\ref{fig:S} illustrates the variation of chemical mixing entropy with temperature. Above 1200K the sublattices are equally occupied and the single-site entropy approximation Eq.~(\ref{eq:S_site}) remains ideal, while a small entropy loss is evident due to the nearest-neighbor pair correlations according to Eq.~(\ref{eq:S_pair}). The entropy drops by approximately $1.5\kB$ by the time the temperature drops to 700K, owing primarily to the segregation of species to distinct sublattices.

\section{Electronic and magnetic properties}

\begin{figure}[h!]
  \includegraphics[width=.48\textwidth]{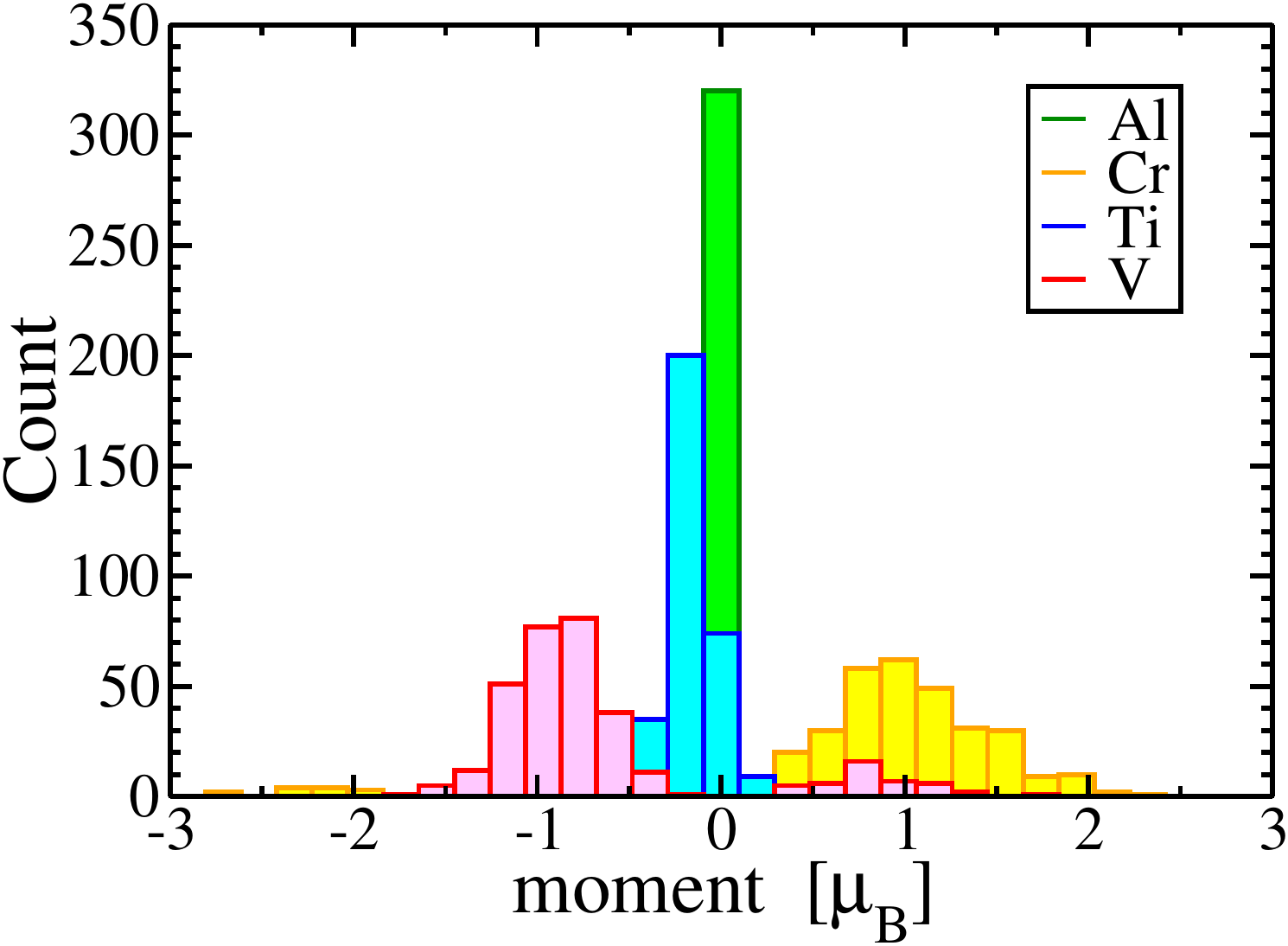}
  \caption{\label{fig:mag} Magnetic moments calculated for cell size $12.8$~\AA~ ($4\times 3.2$~/AA, 128 atoms) at T=1000K.}
\end{figure}

Figs.~\ref{fig:mag} shows the distribution of site-projected magnetic moments accumulated from 10 structures drawn from our simulated ensemble at 1000K. Notice the majority of Cr moments are positive, and the majority of V and Ti moments are negative, achieving a net moment less than 0.02$\mu_B$/atom, demonstrating compensated ferrimagnetism. Al atoms have negligible moment. A small portion of moments point opposite to the majorities for their species, reflecting incomplete chemical ordering, with a fraction of V on the majority Cr sublattice and {\em vice versa}.

\begin{figure}[h!]
  \includegraphics[width=.48\textwidth]{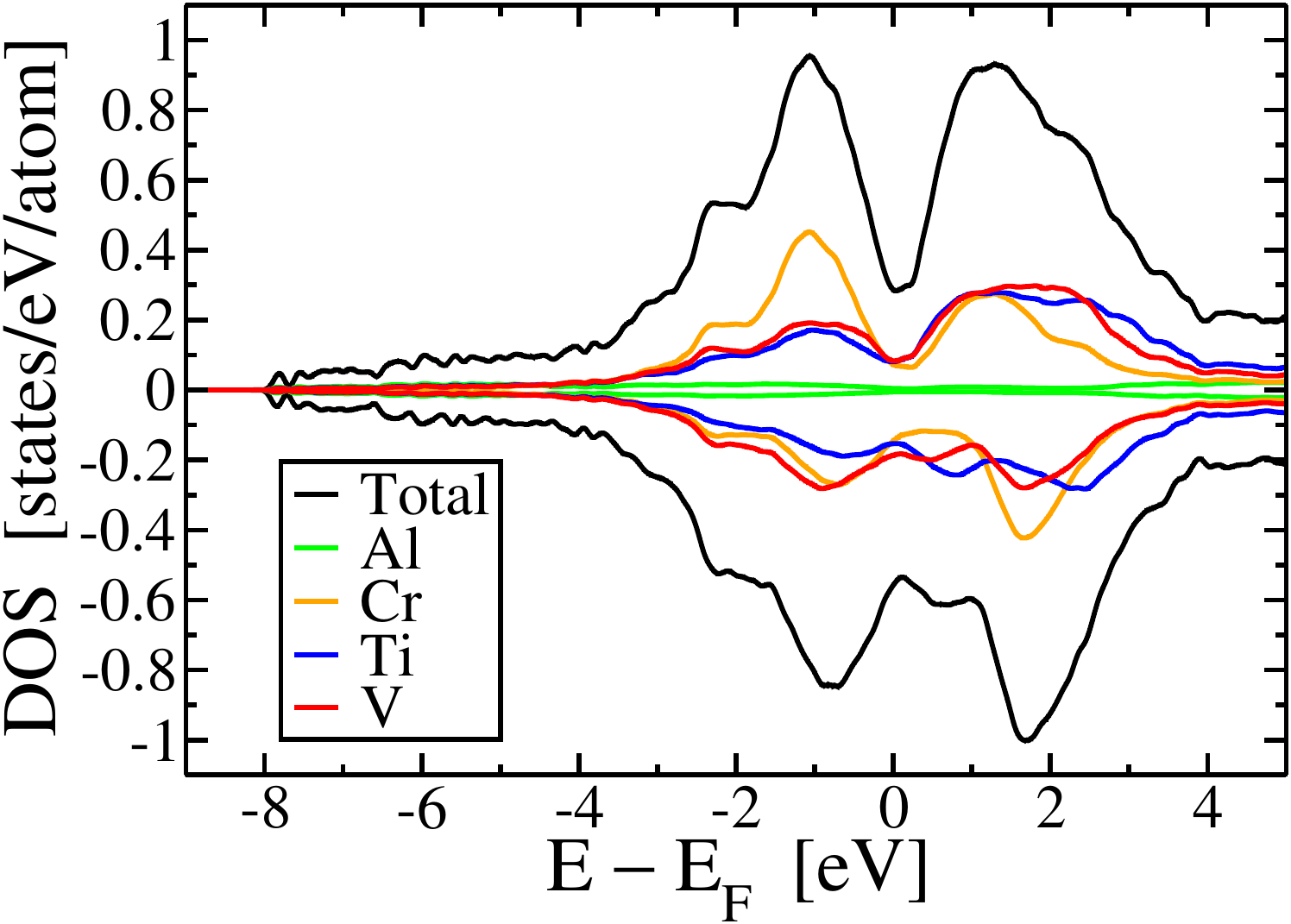}
  \caption{\label{fig:dos} Electronic density of states  calculated for $4\times 4\times 4$ super cells (128 atoms) at T=1000K. }
\end{figure}

Fig.~\ref{fig:dos} shows the species- and spin-projected density of states, averaged over the same 10 structures as discussed above. The Cr atoms exhibit a strong pseudogap in the majority spin (up) states at the Fermi level, and Ti and V also exhibit a pseudogap. A weaker pseudogap is present in the spin down states.

\section{Conclusions}

In summary, we have simulated the high temperature structure of the high entropy alloy AlCrTiV. We find an order-disorder transition occurs around 1200K. The high temperature state is body-centered cubic with short-range chemical order favoring AlTi nearest-neighbor pairs. The low temperature state is B2 with chemical disorder on the sublattices. One sublattice is predominantly (AlCr), the other (TiV), which we denote as B2-(AlCr)-(TiV). Owing to the contrast of neutron scattering lengths, we propose neutron power diffraction experiments as a sensitive test for the presence of this specific ordering. Magnetic neutron diffraction could also be revealing, owing to the opposite polarizations of Cr and V atoms.

Our calculations faced two difficulties. First, when we relax our simulated structures using the PBE-GGA exchange correlation functional, the magnetic moments are largely quenched. To preserve the moments requires holding the lattice constant fixed, close to or slightly greater than the experimental values. Second, calculations using the r2SCAN meta-GGA~\cite{r2SCAN} preserve the magnetism~\cite{Stephen2019PRB} but result in questionable formation enthalpies. Thus the simulated B2-(AlV)-(CrTi) order, and the lower temperature L2$_1$ or Y-III (Al-V-Ti-Cr) structure, may be physically incorrect.

While we favor the PBE-based simulation on the grounds of more plausible the total energies and agreement with established binary alloy phase diagrams, the self-consistent magnetism and the reported presence of Heusler diffraction peaks could favor the r2SCAN-based simulation. For this reason we seek further experimental guidance as to the temperature-dependent chemical ordering, especially utilizing neutron diffraction. We also suggest more detailed calculations of charge and spin transport and their temperature-dependencies as an additional means of resolving these questions.

\section{Acknowledgements}
This work was supported by the Department of Energy under Grant No. DE-SC0014506. This research also used the resources of the National Energy Research Scientific Computing Center (NERSC), a US Department of Energy Office of Science User Facility operated under contract number DE-AC02-05CH11231.

\begin{appendix}
  \renewcommand\thefigure{\thesection.\arabic{figure}}
  \renewcommand\thetable{\thesection.\arabic{table}}

\section{r2SCAN functional}
\label{app:r2scan}
\setcounter{figure}{0}
\setcounter{table}{0}

Calculations using the r2SCAN meta-GGA functional~\cite{r2SCAN,Stephen2019PRB} yielded rather different results compared with the standard PBE GGA functional. On the positive side, self-consistent magnetic solutions were robustly obtained, even for large disordered structures during our replica exchange MC/MD simulations and when subject to positional relaxation. On the downside, r2SCAN failed to properly reproduce the known binary phase diagrams of Al-Cr, Cr-Ti, and Cr-V. Specifically, it fails to predict the stability of AlCr$_2$.tI6, Cr$_2$Ti.cF24, and any CrV structures. Further, it predicted the quaternary Y-III structure to lie on the convex hull (see Table~\ref{tab:dE-scan}), and to be stable against decomposition into pure elements or other potentially competing phases. It seems unlikely Y-III is a genuine ground state, given the disorder reported in experimental synthesis~\cite{Venkat2018,Stephen2019JAP}.

\begin{table}
  \begin{tabular}{lll}
    \hline
    $\Delta E$ & Structure & comment \\
    -62 & Y-III Heusler & Al-V-Ti-Cr ferri insulator\\
    0 & Reference & pure elements \\
    62 & Y-II Heusler & Al-Cr-V-Ti half semimetal\\
    159  & layered tP4 & ferrimagnet \\
    172  & $L=4$ 700K & simulated using r2SCAN\\
    177 & Y-I Heusler & Al-V-Cr-Ti nonmagnetic semimetal\\
    197 & B2-(AlV)-(CrTi) & \\
    198  & simulated & simulated using PBE \\
    206 & B2-(AlCr)-(TiV) & \\
    220 & B2-(AlTi)-(CrV) & \\
    \end{tabular}
    \caption{\label{tab:dE-scan} Energies relative to convex hull of various simulated and hypothetical structures. All compositions are equiatomic. The r2SCAN density functional has been used. Units are meV/atom}
\end{table}

\begin{figure}[h!]
  \includegraphics[width=.48\textwidth]{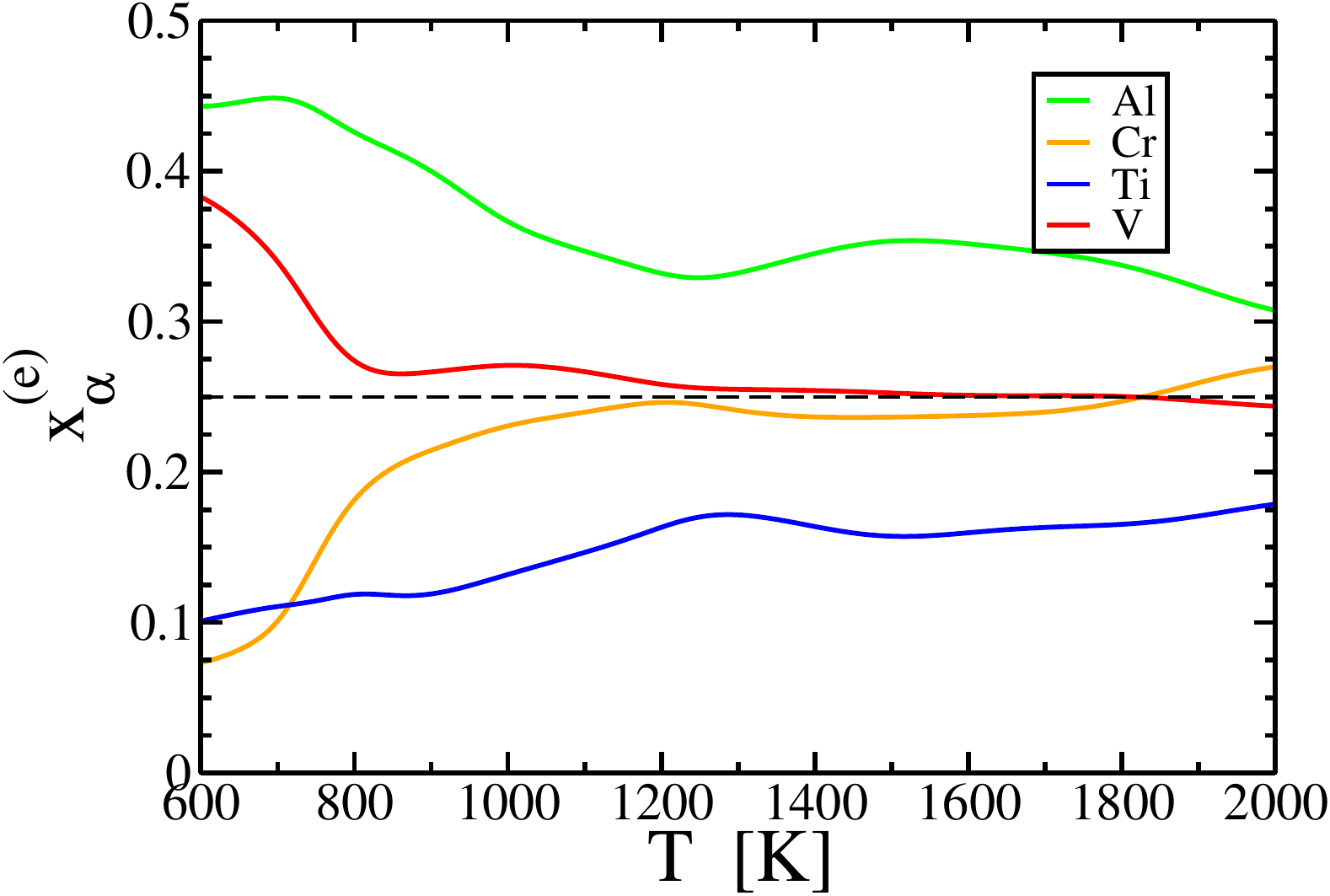}
  \caption{\label{fig:XofT-scan} (solid lines) Temperature-dependent occupation of even (cube vertex) sites. Cell size is $12.4=4\times a=3.1$ (128 atoms).}
\end{figure}

\begin{figure}[h!]
  \includegraphics[width=.48\textwidth]{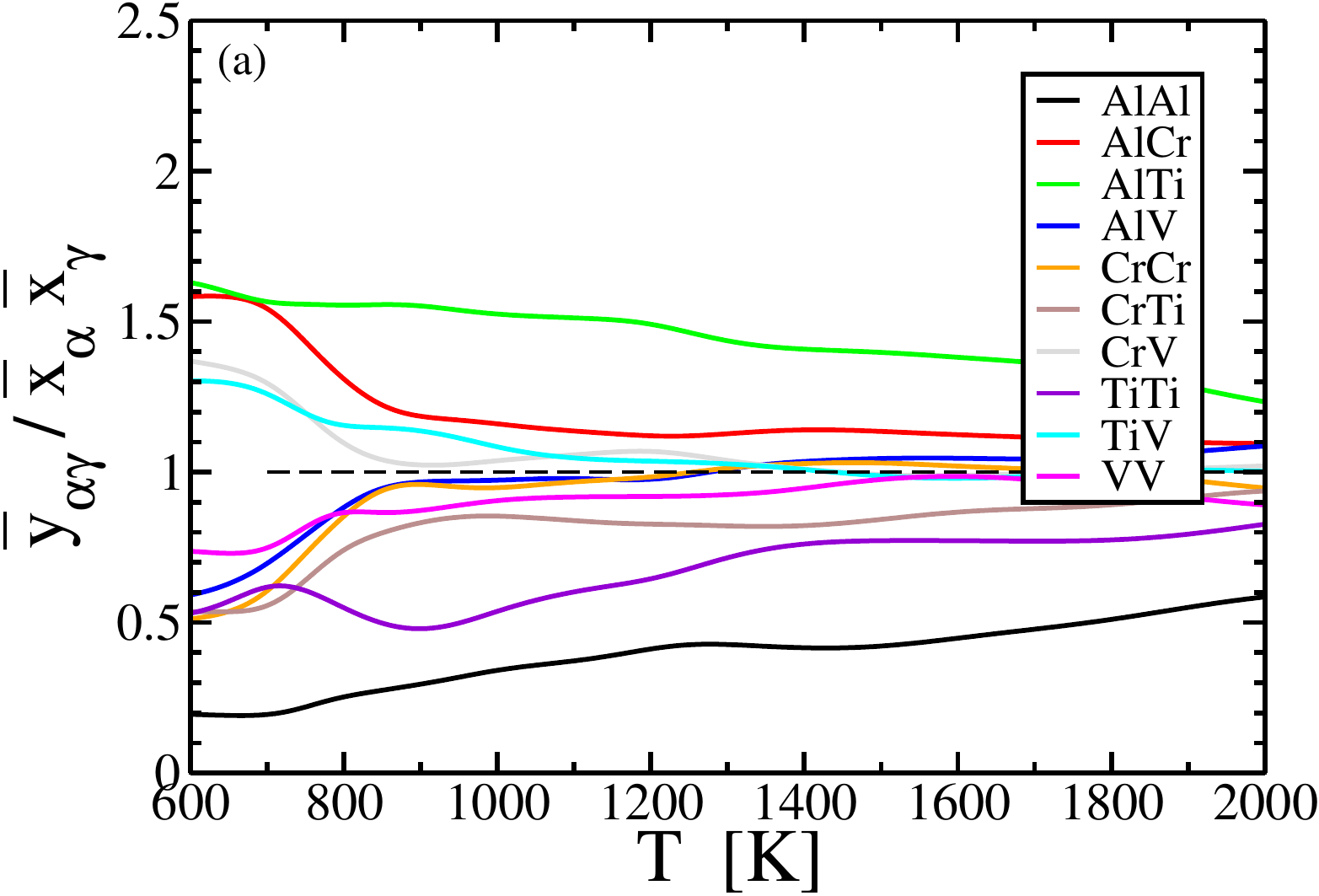}
  \includegraphics[width=.48\textwidth]{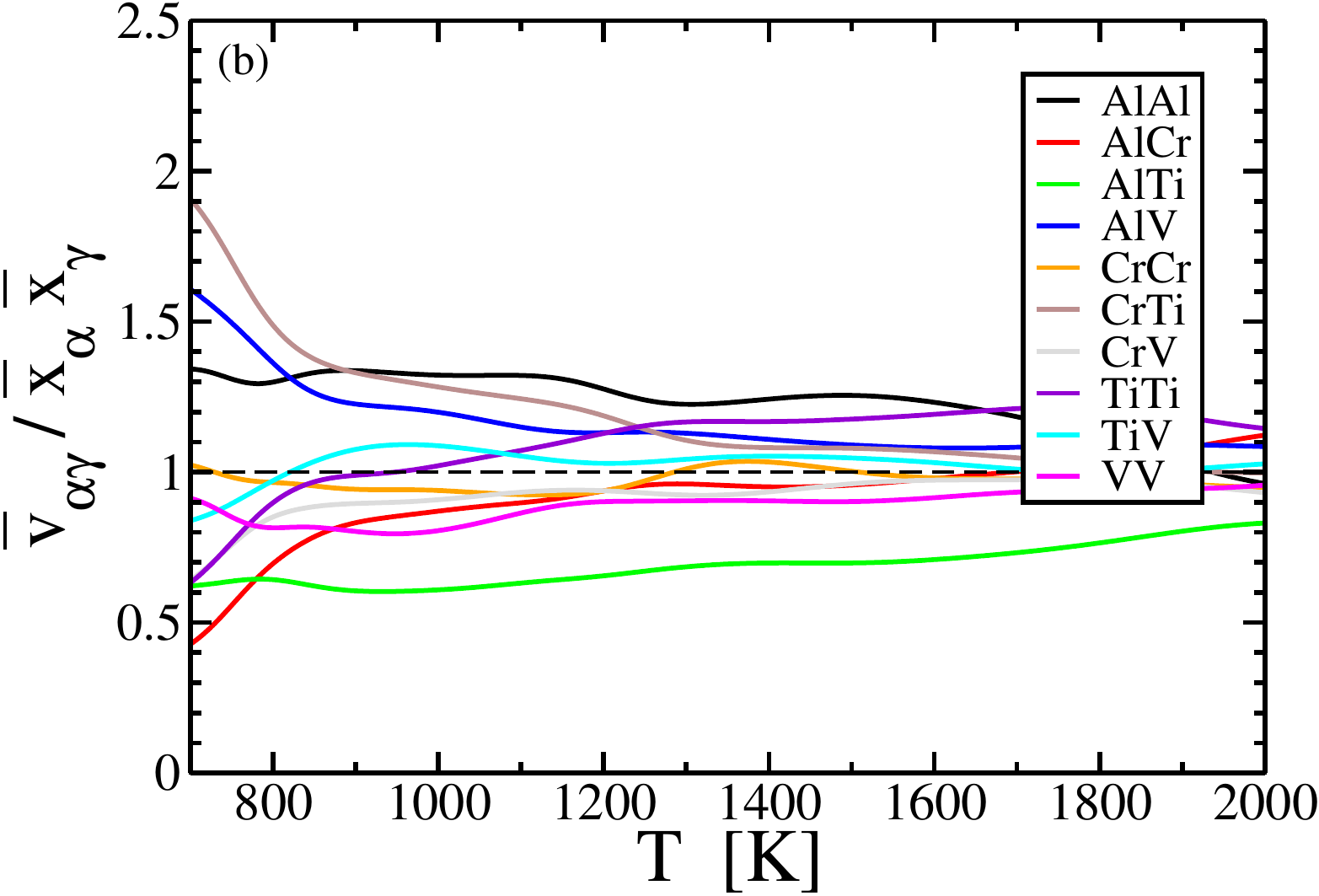}
  \caption{\label{fig:YVofT-scan} Temperature-dependent pair frequencies for (a) nearest neighbors and (b) next nearest neighbors. Cell size is $12.4=4\times a=3.1$ (128 atoms).}
\end{figure}

The temperature-dependent site occupation (Fig.~\ref{fig:XofT-scan}) and pair frequencies (Fig.~\ref{fig:YVofT-scan}) differ from those obtained with the PBE functional (compare with Figs.~\ref{fig:XofT} and~\ref{fig:YVofT}). Preexisting B2-like chemical order alternating Al and Ti on opposite sublattices is present across the entire temperature range and does not diminish appreciably up to 2000K. This may be due in part to the small $L=4$ system size, but it is stronger than we observed using the PBE functional. At high temperature the structure most closely matches the B2-(AlV)-(CrTi) structure. An order-disorder transition does occur, but at the relatively low temperature of 800K, in which Cr and V order relative to Al and Ti, leading to Y-II type Heusler ordering. It will be interesting to study whether this transition is continuous or first order in nature~\cite{Ackermann89}.

\begin{figure}[h!]
  \includegraphics[width=.48\textwidth]{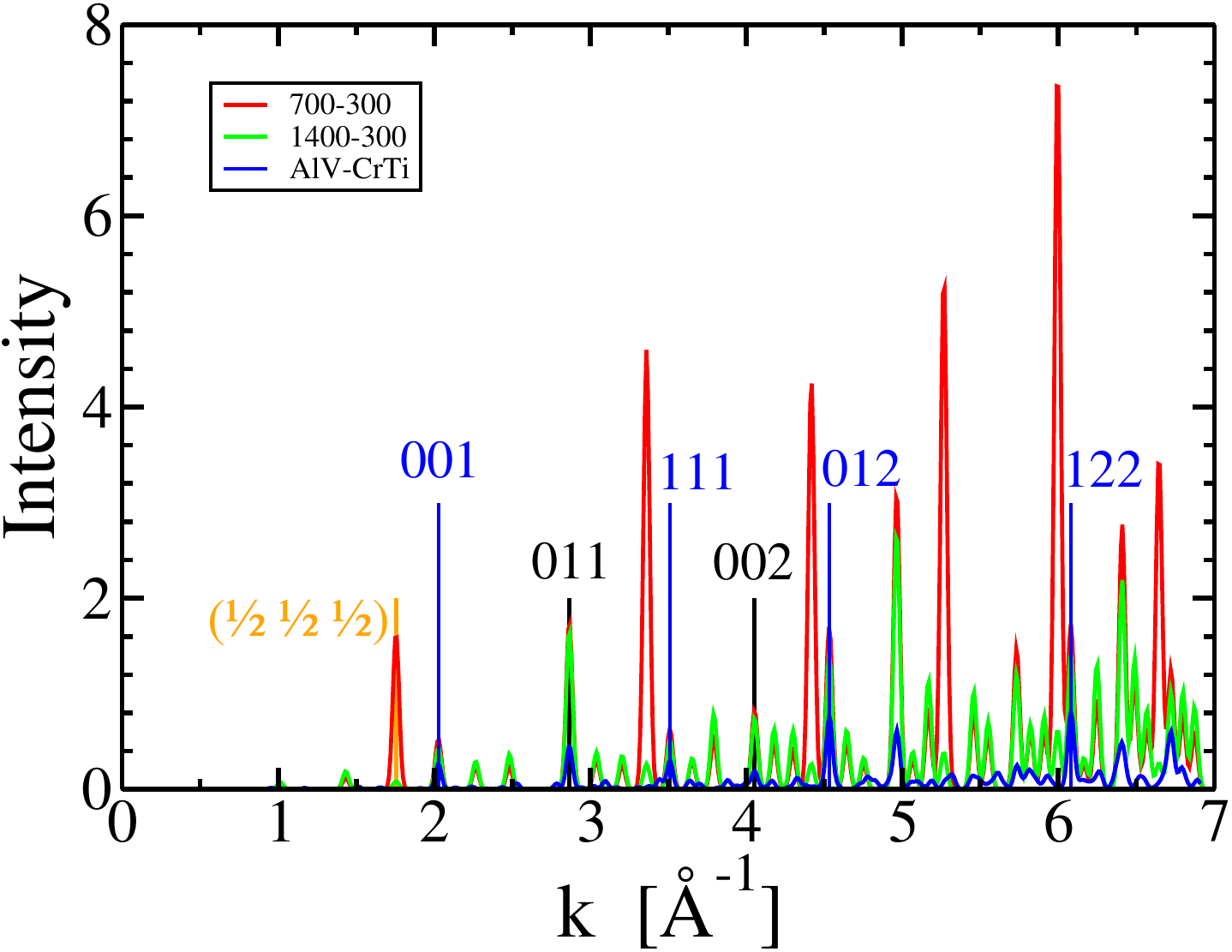}
  \caption{\label{fig:neutron-dp-SCAN} (solid lines) Neutron-weighted diffraction pattern of r2SCAN-simulated structures. Cell size is $12.4=4\times a=3.1$ (128 atoms). Miller indices are marked according to the conventional BCC cell with $a=3.1$~\AA. BCC peaks are marked in black, B2 in blue, and Heusler in orange.}
\end{figure}

Our predicted neutron diffraction patterns are shown in Fig.~\ref{fig:neutron-dp-SCAN}. The emergence of the $(\frac{1}{2}\frac{1}{2}\frac{1}{2})$ peak at low temperature verifies the onset of Heusler ordering. Notably, such a peak was reported in~\cite{Stephen2019JAP}. However, a temperature-dependence was not reported, and it was not asserted that the peak emerged from a majority phase. Rather, the structure was described as majority B2-(AlTi)-(CrV) with an additional L2$_1$ phase, presumable of diagonal (CrV)-Al-(CrV)-Ti, with the chemical order assumed based on the lowest energy Y-III Heusler. Notice that in our predicted neutron patterns, the half-integer Heusler peaks strongly dominate the B2, which in turn dominate the BCC peaks. This again shows the possible advantage of neutron diffraction over X-ray.

\end{appendix}

\bibliography{refs}

\end{document}